\documentclass[11pt]{report}

\usepackage[latin1]{inputenc}
\usepackage{amsmath}
\usepackage{amsfonts}
\usepackage{amssymb}
\usepackage{url}
\usepackage{color}
\usepackage{ifpdf}
\usepackage{wrapfig}
\usepackage{textcomp}
\usepackage{srcltx}
\usepackage{fancyhdr}
\usepackage{setspace}
\usepackage{lscape}
\usepackage{longtable}
\usepackage{url}
\usepackage{subfigure}
\usepackage{graphicx}
\usepackage{multirow}
\usepackage{verbatim}

\definecolor{orange}{rgb}{1.0,0.3,0.0}
\definecolor{violet}{rgb}{0.75,0,1}
\definecolor{darkgreen}{rgb}{0,0.7,0}

\newif\ifdraft
\drafttrue
\ifdraft
 \newcommand{\katznote}[1]{ {\textcolor{cyan}   { ***Dan:   #1 }}}
 \newcommand{\rananote}[1]{ {\textcolor{blue}    { ***Omer:   #1 }}}
 \newcommand{\amnote}[1]{   {\textcolor{magenta} { ***Andre: #1 }}}
 \newcommand{\manishnote}[1]{{\textcolor{orange}    { ***Manish: #1 }}}
 \newcommand{\jonnote}[1]{  {\textcolor{green}     { ***Jon:    #1 }}}
  \newcommand{\note}[1]{ {\textcolor{red}    { #1 }}}
 \newcommand{\jhanote}[1]{  {\textcolor{violet}     { ***Shantenu:    #1 }}}
 \newcommand{\gailnote}[1]{  {\textcolor{darkgreen}      { ***Gail: #1 }}}
 \usepackage[pdftex,colorlinks=true, linkcolor=blue,citecolor=blue,
       urlcolor=blue]{hyperref}
\else
 \newcommand{\katznote}[1]{}
 \newcommand{\rananote}[1]{}
 \newcommand{\amnote}[1]{}
 \newcommand{\manishnote}[1]{}
 \newcommand{\philnote}[1]{}
 \newcommand{\note}[1]{}
 \newcommand{\colenote}[1]{}
 \newcommand{\jonnote}[1]{}
 \newcommand{\jhanote}[1]{}
 \newcommand{\gailnote}[1]{}
\fi

\newcommand{\minihead}[1]{ \vspace*{0.25em} {\noindent \bf #1 } \vspace*{0.25em} \\ }

 \newenvironment{shortlist}{
        \vspace*{-0.8em}
  \begin{itemize}
  \setlength{\itemsep}{-0.3em}
}{
  \end{itemize}
        \vspace*{-0.6em}
}

\newenvironment{changemargin}[2]{
  \begin{list}{}{
    \setlength{\topsep}{0pt}
    \setlength{\leftmargin}{#1}
    \setlength{\rightmargin}{#2}
    \setlength{\listparindent}{\parindent}
    \setlength{\itemindent}{\parindent}
    \setlength{\parsep}{\parskip}
  }
  \item[]}{\end{list}}

\begin{document}

\newif\ifcontribution
\contributiontrue
\ifcontribution
\newcommand{\inputc}[1]{ \input{#1}}
\else
\newcommand{\inputc}[1]{}
\fi

\newif\ifreport
\reporttrue

\newpage

\thispagestyle{empty}

\begin{changemargin}{-2cm}{-2cm}

\begin{center}

{\bf \noindent {\large Survey and Analysis of Production Distributed Computing Infrastructures}

\vspace{2 cm}

\noindent Daniel Katz, University of Chicago \& Argonne National Laboratory

\vspace{0.5 cm}

\noindent Shantenu Jha, Rutgers University

\vspace{0.5 cm}

\noindent Manish Parashar, Rutgers University

\vspace{0.5 cm}

\noindent Omer Rana, Cardiff University

\vspace{0.5 cm}

\noindent Jon Weissman, University of Minnesota

\vspace{2.0 cm}

\noindent CI-TR-7-0811 (Technical Report)

\vspace{0.3 cm}

\noindent 09/16/2011}

\vspace{2.0 cm}

\noindent UNIVERSITY OF CHICAGO

\vspace{0.3 cm}

\noindent Computation Institute

\vspace{0.3 cm}

\noindent 5735 S. Ellis Avenue

\vspace{0.3 cm}

\noindent Chicago, Illinois 60637

\end{center}

\vspace{2 cm}

\noindent \copyright 2011 The University of Chicago

\end{changemargin}

\newpage

\pagenumbering{arabic}

\title{Survey and Analysis of Production\\Distributed Computing Infrastructures\footnote{Please cite as: D. S. Katz, S. Jha, M. Parashar, O. Rana, and J. Weissman. Survey and Analysis of Production Distributed Computing Infrastructures. Technical Report CI-TR-7-0811. Computation Institute, University of Chicago \& Argonne National Laboratory. \url{http://www.ci.uchicago.edu/research/papers/CI-TR-7-0811}}}

\author{Daniel S. Katz, Shantenu Jha, Manish Parashar, \\Omer Rana, and Jon Weissman}

\maketitle

\setcounter{chapter}{1}

\section*{Context}

The material in this report is a draft of a large part of Chapter 3
of ``Abstractions for Distributed Applications and Systems,'' a book
being written by Shantenu Jha, Daniel S. Katz, Manish Parashar, Omer
Rana, and Jon Weissman, to be published by Wiley in 2012.

This report primarily covers production distributed computing
infrastructures that have been used to
develop and deploy large-scale scientific applications. 
We define a production
distributed computing infrastructure as a
set of computational hardware and software, in multiple locations, intended for use
by multiple people who are not the developers of the infrastructure.
We observe
that typically the time scales over which scientific applications are
developed and used is qualitatively larger than the time scales over
which the underlying infrastructure tends to evolve. For instance,
the middleware used and the services and interfaces offered by
many distributed computing infrastructures have changed over recent
years due to changes in providers and other technical, political, and
funding reasons.  Additionally, some of the commercial infrastructures
themselves have developed relatively recently.
However, one component of this landscape has essentially
remained the same: scientific applications and the most commonly used
methods used to develop them. The relatively slow evolution of
scientific applications is both an opportunity and a challenge. It is
a challenge in that once developed, they are hard to modify and adapt
to changes in infrastructure. It is an opportunity in the sense that
if we can design and architect scientific applications correctly they
will be immune to shifts in the underlying infrastructures!

Given the many changes in academic computing infrastructures the world over,
and the fast evolution of commercial infrastructures, this
report is an attempt to provide a topical and focused
analysis of distributed computing infrastructures.
 
The book from which this report has originated provides: (i) a
critical assessment of a number of existing scientific applications
and infrastructures -- to identify gaps between application
requirements and the abstractions and capabilities provided by the
current generation of systems and infrastructure; (ii) a survey of 13
application case studies; (iii) survey of coordination abstractions
and infrastructures currently employed by distributed applications, in
particular identifying mechanisms that may have benefit for future
applications (in addition to those surveyed); and (iv) a survey and
assessment of abstractions and infrastructures within the emerging
area of data intensive applications.  The book is, in part, a
consequence of what we perceive to be a lack of sufficient connection
between: (i) the theory of scientific application development; and
(ii) the theory and practice of deployment over distributed systems.

The method we used to write this report 
was that we asked the following questions:

\begin{enumerate}

\item What is the purpose of your system?

\item What are the main characteristics of your system?

\item What common patterns and usage modes does your system support?

\item What are the common usage modes for applications that use (or will use) your system?

\item How does your system address the usage modes that you have identified?

\item What types of applications and users have been successful in using your system?

\item What are the limitations in the use of your system (i.e. where your system has not been successful)?

\end{enumerate}

\noindent to a set of contributors who were knowledgable about the various
infrastructures (Paul Avery, Henri Bal, Geoffrey
Fox, Wolfgang Gentzsch, Helmut Heller, Adriana Iamnitchi, Scott Lathrop, Hermann
Lederer, Andre Luckow, David Margery, Steven Newhouse, Ruth Pordes,
and David Wallom), and then adapted their responses as the starting
point for the text in sections \ref{sec:pdis}, \ref{sec:rdis}, and
\ref{sec:cdis} of the report.  (Of course, any errors are our
responsibility, not the responsibility of the contributors.)  We then
wrote the other sections of the report to analyze and integrate
these the sections based on contributed
material.

\subsection*{Objectives}

This   report  has two objectives.
First, we describe a
set of the production distributed infrastructures currently available, so
that the reader has a basic understanding of them.  
This includes explaining why each infrastructure was created and made
available and how it has succeeded and failed.
The set is not
complete, but we believe it is representative.  A specific infrastructure
we do not discuss that of the US Department of Energy, because
it isn't really a unified infrastructure in the same sense as those we
do discuss.  Rather, it is a set of independently managed
resources, connected by a high-bandwidth network.

Second, we describe the infrastructures in terms of
their use, which is a combination of how they were designed to be
used and how users have found ways to use them.
Applications are
often designed and
created with specific infrastructures in mind, 
with both an appreciation of the existing capabilities provided by
those infrastructures and an anticipation of their future capabilities.
Here, the infrastructures we discuss
were often designed and created with
specific applications in mind, or at least specific types of applications.
The reader should understand how the interplay between the
infrastructure providers and the users leads to such usages,
which we call usage modalities. These usage modalities are
really abstractions that exist between the infrastructures and the
applications; they influence the infrastructures by representing the
applications, and they influence the applications by representing the
infrastructures.

\subsection*{Motivation}

To analyze why an infrastructure was put together and made
available, we need to understand the overall design decisions and
design considerations.  We know that these are driven by several 
factors, including politics and funding, expectations of which applications will be run
on the infrastructure and of who the users will be, and  the desire of the infrastructure providers
to try out new technologies.

To describe how an infrastructure is used, we consider its usage
modes. These can be described as combinations of a set of modalities
(based on those previously published in~\cite{usage_modalities}):

\begin{itemize}
\item User intent: production, exploration/porting, education
\item When to run: batch (normal); interactive (when the user is ready); urgent (immediate); urgent (not immediate, but high priority); reservation (at a set time)
\item Submission mechanism: command line; grid tools; science gateway; metascheduler (automatically selected)
\item Targeted resources: use of multiple resources of the same type\footnote{Type here is used to mean HPC compute, HTC compute, storage, visualization, etc.} within the infrastructure; use of multiple types of resources within the infrastructure; Coupling of these resources with other resources that are not part of the infrastructure
\item Job/resource coupling: independent; independent but related (e.g., ensemble); tightly coupled (e.g., must be coscheduled with low-latency, high-bandwidth network connection); dependent (e.g., workflow)
\end{itemize}

For example, one usage mode could be when a user runs an MPIg\footnote{MPIg~\cite{mpig} is a tool that allows one to run an MPI
application across more than one system.}
application,
as part of a set of production runs, using a reservation,
submitted through grid tools, on a pair of HPC systems, where the two
applications are tightly coupled.  Another example might involve a user
running a production workflow for a forecast hurricane, using urgent
scheduling, submitted through a metascheduler, targeting multiple HPC
resources and storage resources, with dependent coupling between jobs.

\subsection*{Overview}

Many production distributed-computing infrastructures are now available.  These
can be classified into three categories: science, research, and commercial.
TeraGrid (now transitioned into XSEDE) and DEISA are two roughly similar
science infrastructures, the former based in the 
US and the latter in Europe. Each is intended to ``unify''
activities involving multiple large-scale parallel systems across the
geographical area it covers.  OSG, EGEE (now transitioned into EGI),
and NGS are roughly similar science infrastructures that
are more oriented to high-throughput computing, in the United
States, Europe, and the United Kingdom, respectively.  All five of
these science infrastructures are primarily intended to be used to
achieve research results in application science.  Grid'5000, in France,
and DAS, in the Netherlands, are research infrastructures aimed more
at computer science research.  PlanetLab is a worldwide research
infrastructure aimed at computer science research, and FutureGrid is an
emerging experimental testbed that will transition into being part of
the US national cyberinfrastructure. The commercial Amazon Web Services
and Microsoft Azure infrastructures are a mixture of commercial usage,
science, and research.  From the
points of view of Amazon and Microsoft, these infrastructures are products
that support their company. Unlike
the science infrastructures, they are not open, meaning that users
cannot easily interact with the infrastructure providers to ask for new
features.

The sections of this report describe a number of 
science, research, and commercial infrastructures,
prior to a discussion and comparison of
the various infrastructures.
Each infrastructure description in the next three sections is laid out as
follows: an introduction to the infrastructure, generally including history,
source of funding, mission and vision, management, and a roadmap of
where the infrastructure is going; the characteristics of the
infrastructure, often including the resource provisioning or
aggregation model; the patterns and usage modes employed in the
infrastructure; and the successes and limitations of the infrastructure.
Please note that the infrastructures described were chosen as
representative of the infrastructure landscape
at the time of writing, and we recognize that these infrastructures
are quite disparate in goals, scope, scale, and
targeted user communities.

\subsubsection{Issues related to the timing of this report}

Most of this report was completed at the end of 2010, with some additions made in mid-2011.  It provides a snapshot
of the state of the infrastructures discussed and gives an outline of where
we think the infrastructures are heading, based on discussions,
our own knowledge, and assorted public material.
During the writing of this  report,  EGEE transitioned into EGI, 
TeraGrid transitioned into XSEDE, and Open Science Grid will transition into a new program.
Infrastructures are always changing.

\section{Science Production Distributed Infrastructures\label{sec:pdis}}

In this section, we discuss five national and international science production
 distributed infrastructures.

 \subsection{TeraGrid} 

Funded by the US National Science Foundation (NSF), TeraGrid~\cite{teragrid,TeraGridScience} was an advanced, nationally distributed, open cyberinfrastructure that enabled and supported leading-edge scientific discovery and promoted science and technology education.
TeraGrid included resources (supercomputers, experimental, storage, and visualization systems, data collections, and science gateways) connected by high-bandwidth networks and integrated by software and by coordinated policies and operations, all supported by computational leaders and technology experts. 
At the end of the TeraGrid project (June 2011), TeraGrid resources included more than 2 petaflops of computing capability and more than 60 petabytes of online and archival data storage, with rapid access and retrieval over high-performance networks. Researchers could also access more than 100 discipline-specific databases.

\minihead{History}
In 2001 NSF made an award to four centers to establish a distributed terascale facility (DTF). The DTF became known to users as TeraGrid, a multiyear effort to build and deploy the world's largest, fastest, most comprehensive distributed infrastructure for general scientific research.

The initial TeraGrid 
was
homogeneous and
``griddy,'' with users foreseen to be running on multiple
systems, both because their codes could run ``anywhere'' and
because, in some cases, multiple systems would be needed to support
the large runs that were desired. 
TeraGrid included a set of software packages that were identical on all systems.
TeraGrid subsequently expanded in capability and number of resource providers.
The expansion introduced heterogeneity and thus added complexity to the grid ideals of the initial DTF, since the common software no longer could be identical.  This situation led to the concept of common interfaces, with potentially different software underneath the interfaces.  
Additionally, the users of the national centers' supercomputers
were merged into TeraGrid, motivating TeraGrid to increase its focus on supporting these users and their traditional parallel/batch usage modes.

\minihead{Mission/Vision}
The TeraGrid mission was twofold: (1)  to enable and support leading-edge computational research through the provision of an advanced, distributed, comprehensive, and open cyberinfrastructure and (2) to promote the use of this cyberinfrastructure in research and education.

TeraGrid achieved its purpose and fulfilled its three goals: 
\begin{itemize}
\item {\bf Deep}: let the most experienced users use the most powerful computational resources and advanced computational expertise/support to do their work;
\item {\bf Wide}: find larger and more diverse communities of researchers and educators who can use the resources, including through science gateways; 
\item {\bf Open}: facilitate simple migration between TeraGrid and other resources through use of open interfaces, partnerships with other grids, and collaborations with research and education institutions.
\end{itemize}

\minihead{Management}
TeraGrid's Grid Infrastructure Group (GIG) at the University of Chicago,
working in partnership with eleven resource provider sites, coordinated the functions and operation of TeraGrid~\cite{teragrid}.

\minihead{Roadmap/Future}
 The overall management of the TeraGrid changed in 2011, as the TeraGrid transitioned under a new NSF funding program called eXtreme Digital, or XD.  The XD solicitation called for broadening access as the main new feature, and the new project that has replaced TeraGrid
 is called XSEDE.  XSEDE intends to continue many of the successful parts of TeraGrid, and in general the features described here are intended to describe XSEDE as well as TeraGrid, unless otherwise mentioned. 
SInce NSF awards to providers last for two to four years,
TeraGrid resources and resource providers often changed;
and this process is also expected to continue with XSEDE.

\subsubsection{Characteristics}

The nodes of TeraGrid spanned a wide variety of architectures, sizes, and purposes: 
clusters,
massively parallel systems, 
shared-memory systems, and
systems dedicated to remote visualization, ranging from entry-level and experimental resources to a 1-petaflop system.
The TeraGrid network provided high-capacity links among these resources. Each resource provider maintained at least 10 Gbps of connectivity to one of three TeraGrid hubs,
which
were interconnected via 10-Gbps fiber optic links.
In 2009, TeraGrid delivered about 700 million core-hours to about 4,800 users.

TeraGrid had a single allocations process with a  national peer review; a single point of access via a user portal; a set of coordinated software and services kits based on GT4 technology deployed on each resource according to its architecture and purpose; and a unified user support, documentation, training, and educational system.

The TeraGrid project  introduced important new methods and tools,
such as science gateways~\cite{TGSG},
for making high-end computing available and useful to a wide range of academic communities.
The Campus Champions~\cite{TGCC} program  actively spread news on campuses across the country about the availability of resources for research and education. 

\subsubsection{Usage Modes}

TeraGrid users submitted jobs to the batch queues of the particular system on which they wanted to run their application, either directly from that system or indirectly from another system, using Grid software.  Users were encouraged to use the TeraGrid User Portal to monitor the batch queues and to use the batch queue predictor to assist them in selecting the systems best suited to their needs. Users could request special handling of jobs, including access to dedicated system time, to address special job-processing requirements.

TeraGrid usage modes, as shown in Table~\ref{teragridusage}, can be divided in deep and wide categories, two of the three TeraGrid goals.  Note that this table shows numbers of users, not the amount of usage.  Deep users use far more of the resources, both per user and in sum, than do the wide users.  In fact, in the third quarter of 2009, the top 24\% of the users used more than 80\% of the resources.

The deep usage modes of TeraGrid resources, by experienced computational scientists and engineers, exploited TeraGrid's large-scale resources and the intellectual expertise of the staff at the resource providers. Included was  the ability to run batch jobs on the high-end resources, as well as data storage, management, analysis, and transfer capabilities. Complex and heterogeneous work and data flows, urgent computing, and interactive computing are also being enabled. Moreover, as new methodologies for large-scale, data-intensive computational science (data mining, statistical analysis, etc.) continue to explode in popularity and importance, TeraGrid/XSEDE must support the high-end users in these modalities also.

The wide usage modes of TeraGrid aimed to increase the overall impact of TeraGrid's advanced computational resources on larger and more diverse communities through user interfaces, domain-specific portals, and enhanced support that facilitate scientific discovery without requiring people to become high-performance computing experts. 
Features included the development and support of simpler and more powerful interfaces---ranging from common user environments to science gateways and portals, through more focused outreach and collaboration with science domain research groups---and educational and outreach efforts that will help inspire and educate future scientists.

\begin{table} \footnotesize
\begin{tabular}{|l|l|l|} \hline
    {\bf Usage Mode} & {\bf Type} & {\bf Number of Users}\\
    \hline \hline
    Batch Computing on Individual Resource  & mostly deep & 850 \\
    \hline
    Exploratory and Application Porting & N/A & 650 \\
    \hline
    Science Gateway Access & mostly wide & 500 \\
    \hline
    Workflow, Ensemble and Parameter Sweep & deep \& wide & 250\\
    \hline
    Remote Interactive Steering and Visualization & mostly deep & 35 \\
    \hline
    Tight-Coupled Distributed Computation & deep & 10 \\
    \hline \end{tabular}
\caption{\em TeraGrid usage mode distribution for 2006, the latest year for which data is available. \label{teragridusage}}
  \normalsize
\end{table}

Additionally, within the open usage modes, TeraGrid wanted to enable, simplify, and encourage scaling into its large-scale resources. To this end, TeraGrid provided interfaces and APIs, and it went further to include appropriate policies, support, training, and community building.
TeraGrid tried, with varying levels of success, to make
its cyberinfrastructure accessible from, and even integrated with, cyberinfrastructure of all scales, including not just other grids but also campus cyberinfrastructures and even individual researcher laboratories and systems.

Numerous commercial and academic applications were available across the various computing systems to support users from multiple domains.  More than 75 software applications supported research across multiple domains, including
molecular biosciences, chemistry, physics, astronomy, materials research, chemical and thermal systems, and atmospheric sciences.
TeraGrid conducted surveys and interviews with the user community throughout the year to assess their needs and requirements, and it utilized this information to improve the resources and services offered to the user community.  This process will be more formal in XSEDE.

\subsubsection{Successes and Limitations}

The success of TeraGrid was attested to by the impressive number of
publications resulting from its use.
Indeed, each year, the TeraGrid research community reported results in over 1,000 publications in various professional journals.

TeraGrid evolved far beyond the scope and the architectural stages adopted in 2001 and 2005. As called for in the NSF ``eXtreme Digital'' (XD) solicitation,
a new technological and organizational framework was needed, and XSEDE intends to provide this.

TeraGrid's mission evolved from an infrastructure that supports distributed execution across multiple sites to a collection of mostly stand-alone HPC machines.
A common complaint about TeraGrid was that it supported the use of individual resources well but did not focus on the challenge of collectively utilization of multiple machines. 
In other words, TeraGrid addressed the requirements to enable applications to scale up well, but did not address the requirements to scale out as much~\cite{dpa-grid2009}.  It is unclear how this will change in XSEDE.

 \subsection{DEISA} 

Resources from a distributed set of European HPC centers are integrated in the Distributed European Infrastructure for Supercomputing Applications (DEISA) \cite{deisa} to provide a common set of services for HPC users primarily in Europe.

\minihead{History}
The DEISA Consortium deployed and operated DEISA, cofunded through the EU FP6 DEISA project, from 2004 to 2008. The consortium has continued to support and further develop the distributed high-performance computing infrastructure and its services through the EU FP7 DEISA2 project with funds for another three years until 2011. 

\minihead{Mission/Vision}
The mission of DEISA is to support European scientists through an integrated and unified infrastructure with remote, user-friendly, secure access to a European HPC service to solve big-science (grand challenge) problems.

\minihead{Management}
DEISA supports and enhances activities and services relevant to enabling applications, operations, and technologies, as these are needed to effectively support computational sciences in the HPC area.  The DEISA Extreme Computing Initiative (DECI), launched in 2005, has regularly supported grand challenge projects to enhance DEISA's impact on the advancement of computational sciences. By selecting the most appropriate supercomputer architectures for each project, DEISA has opened up the most powerful HPC architectures available in Europe for the most challenging projects. This service provisioning model has been extended from single-project support to supporting virtual European communities. Collaborative activities have been carried out with new European and other international initiatives. 

\minihead{Roadmap/Future}
Of strategic importance has been the cooperation with PRACE, the Partnership for Advanced Computing in Europe~\cite{prace}. PRACE has first prepared for the installation of a limited number of leadership-class Tier-0 supercomputers in Europe and is now building an ecosystem of Tier-0 and national Tier-1 resources.\footnote{In Europe, Tier-1 is the term used for national centers, and Tier-0 is the term used for pan-European centers.  The use of these terms implies a pyramid, with a small number of
Tier-0 centers at the top, a larger number of Tier-1 centers below, and possibly more tiers below that.} The key role and aim of DEISA has been to deliver a turnkey operational solution for such a persistent European HPC service, as suggested by ESFRI, the European Strategy Forum on Research Infrastructures (a strategic instrument to develop the scientific integration of Europe and to strengthen its international outreach.)

\subsubsection{Characteristics}

DEISA has operated on top of national services. It includes the most powerful supercomputers in Europe with an aggregated peak performance of over 2 petaflops in 2010. The supercomputers are interconnected with a dedicated 10-Gbps network, based on GEANT2 and the National Research and Education Networks.  DEISA has operated a high-performance global file system, facilitating data management and community access to data repositories. Core DEISA services include single sign-on based on common authorization, authentication, and accounting; the provision and maintenance of the DEISA Common Production Environment (DCPE); and various middleware stacks.

As a principle, all DEISA partners provide about
5\% of their national HPC resources for DEISA projects.
In 2008, 66 proposals were submitted to DECI, from 15 European countries,
involving co-investigators from North and South America, Asia, and Australia. 
A total of 134 million normalized CPU-hours\footnote{DEISA normalizes
CPU-hours so that resource requirements can be compared with systems
with CPUs of varying capability.  DEISA has chosen to use an IBM P4+
CPU-hour as its normalized unit.} were requested.  Of these, 42 proposals
were accepted, using 48 million normalized CPU-hours. In addition, 8 million CPU-hours were awarded to science communities.  These projects were executed in 2009.
In the next DECI call (DECI-5 for 2010 access), 69 million CPU-hours were awarded to 50 projects; and in DECI-6 (for access in 2010 and 2011), 91 million CPU-hours were awarded to 56 projects, in addition to another 12 million CPU-hours awarded to science communities. 

\subsubsection{Usage Modes}

The DEISA infrastructure essentially supports large, single-site capability computing through highly parallel batch jobs. Proposals for grand challenge computational projects are peer reviewed for scientific excellence, innovation potential, international aspects, and national preferences.  The best suited, and, when required, most powerful supercomputer architectures are selected for each project. DEISA also supports multisite supercomputing for many independent supercomputer jobs (e.g., parameter sweeps) through various technical means (e.g., UNICORE~\cite{unicore}, DESHL, Globus~\cite{globus}, Application Hosting Environment~\cite{ahe}), using the DEISA global file system and its single name space. Data management is also supported via GridFTP.

DEISA supports mainly four application usage modes: single-job parallel programs for efficient usage of thousands of processor-cores (including ensembles, namely, multiple copies of one application with different input parameters), data-intensive applications with distributed file system support, workflow applications to combine several compute tasks (simulation, pre- and post-processing steps), and coupled applications.
 
The DEISA system addresses these modes by job management and data management services developed with the distributed nature of the system in mind.  The job management service is realized by a user interface for submitting jobs and workflows to distributed compute systems. Alternatively, users can log in to a target system and submit jobs directly, which is what is done for the vast majority of DEISA jobs. 
Workflow management, currently based on UNICORE, enables the coordinated execution of multiple interdependent subjobs running on various platforms. 

The data management service has been based on IBM's Global Parallel File System (GPFS). DEISA members provide access to this DEISA-wide shared file system to enable users to access their data transparently from every partner site. In addition, for systems not capable of attaching to GPFS directly, GridFTP, a component of the Globus Toolkit, is used to transfer data.

\subsubsection{Successes and Limitations}

DEISA has had three strong successes.  First, it has created a unified infra\-structure for accessing the most powerful European HPC systems, using grid middleware such as Globus and UNICORE (single sign-on). The same middleware is also being used in EGI's HTC systems (see~\S\ref{egee}), and it thus allows users to satisfy their growing computer needs from HTC to HPC without having to change their access methods.
Second, through DEISA, the consequences of Moore's law have been mitigated for many countries with only one or no national supercomputer center, since  a supercomputer at the end of its productive lifetime after some five years is hardly still usable for leading-edge computational science projects.
Third, DECI has proven to be successful, and a large amount of science has been done~\cite{DEISA_DIGEST_2010}.

Additionally, DEISA has been a single contact point for supercomputer time allocation all over Europe, which simplifies proposals for users and allows centers to optimize computer time usage (aiming to direct projects to the ``best suited execution site''). This is a success, as it gives access to HPC resources to researchers who would otherwise not have access to these computers, but it is also a limitation in some sense, as it makes using multiple systems or metascheduling applications for best time to solution difficult. 

Another limitation of DEISA is that there is no coscheduling service (also, no advance reservations) for all sites.  Some tools (e.g., HARC~\cite{harc}) have been evaluated, but none are widely deployed in DEISA. This is not a technical problem but has to do with the way the HPC resources are used, namely, for rather long-running, large jobs (which is different from HTC resources). Furthermore, HPC resources are often overbooked (i.e., loaded to close to 100\%); using advance reservation would cause lost time by having to block resources to satisfy the advance reservation. This is not as problematic in an HTC setting where small jobs can be used for backfilling.

The limited usage of the component resources as part of DEISA also is a problem. The hardware resources for the European supercomputing infrastructure are funded from national budgets of the member states; Europe does not provide central funding for supercomputers and or ensure persistence for a European HPC infrastructure. Therefore, DEISA includes only a fraction of these nationally funded resources.

 \subsection{Open Science Grid} 

Open Science Grid (OSG)~\cite{osg} is a US distributed-computing infrastructure for large-scale scientific research, primarily loosely coupled, high-throughput computing.  OSG contributes to the Worldwide Large Hadron Collider (LHC) Computing Grid as the US shared distributed-computing facility used by the ATLAS and CMS experiments.  OSG collaborated with the EGEE project in Europe to provide interoperating federated infrastructures that could be used transparently by the LHC experiments' software.

\minihead{History}
OSG began in 2005, with many
roots that started before 2000, including the needs for computing from
the Laser Interferometer Gravitational Wave Observatory (LIGO)~\cite{ligo} and US LHC~\cite{lhc} projects and three
computer science/physics projects (PPDG, GriPhyN, and iVDGL) that came together as
the Trillium and then Grid3 projects~\cite{avery}.

\minihead{Mission/Vision}
OSG, jointly funded by the US Department of Energy and NSF, is an open collaboration of scientific, research and educational communities, including users and hardware and software providers, to build, operate, use, and evolve a shared, national high-throughput computational facility based on common concepts, technologies, and processes. 
OSG staff maintain the distributed computational
facility; provide support for the facility's (over 3,000) users,
software, and services; and manage the interfaces to external
contributors and peer grid infrastructures. 

\minihead{Management}
OSG is defined by the set of operational services, software, and
processes that enable the contributed resources to act as a coherent
distributed system in support of the users. OSG extends the
capabilities and capacities of the facility and enables and interfaces to
campus and regional cyberinfrastructures.
OSG does not own or
maintain computer resources; these are maintained by the owners.
Nor does OSG develop software (middleware and applications); these are
acquired from external software development groups. OSG
does support integrated software releases (based on the OSG's
Virtual Data Toolkit (VDT)~\cite{VDT}, which includes Condor-G~\cite{condor-g}, Globus~\cite{globus}, VOMS~\cite{voms}, etc.) and works closely with
software developers to ensure the current and future needs of the user
communities will be met.

\minihead{Roadmap/Future}
The OSG facility expands continuously as a result  of the integration of new
sites, the installation of new resources, and the joining of new member
communities together with new partnerships and
collaborations.  OSG's current funding is expected to end in mid-2011, and
the OSG management team has proposed a new OSG to follow.

\subsubsection{Characteristics of OSG}

OSG is a
sustained collaboration of domain
researchers,
computer scientists,
computing resource administrators, software providers, and OSG staff. The time and effort
needed to maintain this organization and manage the work
are significant and
receive ongoing attention. It took over a year to
define consortium governance; and
as the
organization matures, the details have been revisited about every two years.

OSG usage includes physics event simulation, molecular dynamics,
protein structure prediction, biology, climate, text mining, and
computer science. The user communities with the most
challenging needs are the large physics collaborations in the US. The
OSG provides the US computing infrastructure for 
the LHC ATLAS~\cite{atlas} and CMS~\cite{cms}
experiments. Other major users are
LIGO~\cite{ligo}, the Tevatron experiments
(D0 and CDF)~\cite{d0}, and the STAR Relativistic Heavy Ion
Experiment~\cite{star}. This diverse mix of (currently) more than 30 user
communities and applications ensures the evolution of a generic,
nationwide cyberinfrastructure, currently including more than 60
sites.  The late-2010 average daily use of OSG was more than 45,000
CPU-days per day. The physics
communities account for about 85\% of the usage. 

OSG's methods and processes are
based on virtual organizations (VOs), ranging from dynamic, ad hoc
collections with a specific short-term purpose to long-lived, stable
collaborations with well-defined governances. VOs can contain other VOs; 
can interface with each other and share resources; and 
can have common services, common organizational policies and methods,
and common members. 
Where communities layer their own grid over OSG, 
the community's VO registers with OSG to
enable members of the user community to use additional OSG resources
and services. 
This approach enables
university, regional, research, and scientific communities with their
own grid infrastructures to integrate with and/or rely on some or all
of the OSG facility. OSG itself is a VO with people, resources,
and services having a common purpose and governance.

OSG provides access to and sharing of resources through common, shared services for
monitoring, accounting, security, problem reporting, and tracking.
Additionally, OSG provides a common, shared integration
and validation facility and processes 
for
testing new releases of
software, services, and applications.

OSG packages, releases, documents, and supports a well-defined set of software to enable the interfaces to and use of the
contributed resources. This software, including 
the VDT, provides technologies used by both OSG and other
infrastructures, such as TeraGrid and EGEE. Each project, including OSG,
augments the VDT with specific configuration scripts and utilities for
its own environment and users. 

OSG works to bridge its
infrastructure and services with other grids, enabling transparent access, movement of jobs, and management of data.
This strategy is crucial for the
main OSG stakeholders, such as LHC scientists for whom OSG is
``merely'' the US part of the larger worldwide Worldwide LHC Computing Grid (WLCG).

\begin{table}
\begin{tabular}{|l|}
\hline
\begin{minipage}{11.5 cm}
\vspace{0.1cm}

Key Principles

\begin{shortlist}

\item Phased deployment with clear operations model.

\item OSG technologies and user needs allow for 100\% use of all resources.

\item Services should work toward minimizing their impact on the
hosting resource, while fulfilling their functions.

\item Local rules come first: All services should support the ability
to function and operate in the local environment when disconnected
from the OSG environment.

\item Supplementary services: OSG will provide baseline services and a
reference implementation. Use of other services will be allowed. VOs
can deploy additional services.

\item Middleman: Users are not required to interact directly with
resource providers. Users and consumers (possible programs) will interact
with the infrastructure and services.

\item Inclusive participation: The requirements for participating in
OSG should promote inclusive participation both
horizontally (across a wide variety of scientific disciplines) and
vertically (from small organizations such as high schools to large ones
such as national labs).

\end{shortlist}

Best Practices

\begin{shortlist}

\item OSG's architecture is VO-based.  Most
services are instantiated in the context of a VO. OSG's baseline
services and reference implementation can support operations within
and shared across multiple VOs.

\item Resource providers should provide the same interface to
local use of the resource as they do to use by the distributed
services.

\item Every service will maintain state sufficient to explain expected
errors. There will be methods to extract this state. There will be a
method to determine whether the service is up and usable.

\item OSG's infrastructure will support development and execution of
(user) applications in a local context, without an active connection
to the distributed services.

\end{shortlist}

\vspace{0.1cm}
\end{minipage}
\\ \hline
\end{tabular}
\caption{Some key principles and best practices of OSG (paraphrased).\label{Tab:OSGPrinciplesPractices}}
\end{table}

Another characteristic of OSG is the set of underlying
principles, listed in Table~\ref{Tab:OSGPrinciplesPractices}.
In all OSG activities, these principles are applied to the implementation
concepts and design and are measured against the practices and
procedures. This approach contributes to a coherent, consistent technical
path through a diverse set of developments.

\subsubsection{Usage Modes}

OSG offers a
data center service relationship to its users as customers,
including
the standing operations, support, and organizational
services that a user community can depend on and
use with little overhead. The modes of use include ``guaranteed'' (where
the resources are owned by the user community), ``agreed upon
expectations'' (where there has been negotiation between the user and
resource owner communities on the expected level of throughput and
support), and ``opportunistic'' (where the users make use of available
resources based on the standard policies of the owners as members in
the OSG Consortium).

OSG helps integrate and support the use of multiple infrastructures as
needed by its members, through multiplexing software and services that
hide differences in infrastructure, as well as bridges and gateways
that transform and translate information and control to the interfaces
and schema of the differing services
of the production infrastructure and the resources accessible through
it. Some of these services are defined as ``critical'' to the use of
the infrastructure by one or more of the user communities. For
example, the US LHC relies on the publishing of information about OSG
resources to the WLCG. The availability of such services is measured,
with the target availability being agreed to with the users. Critical
services (e.g., the information publisher) are being made 
available. 

OSG is particularly effective
for
high-throughput, pleasingly parallel\footnote{We 
prefer the term ``pleasingly parallel'' to the somewhat more common
``embarrassingly parallel,'' since we don't find parallelism to
be at all embarrassing.}
applications;
job runs of between one hour and several days;
jobs that can be checkpointed;
explicit management of large scale data movement and storage; and
ensembles that can effectively run across a large number of resources.
Table~\ref{Tab:OSG_apps} summarizes the types and
characteristics of applications running on OSG. Any 
application may have one or more such characteristics. 

Applications are supported by OSG software, which provides
capabilities for remote job scheduling, resource selection, and data
movement and access. Particular aspects of support for the
different application types are shown in Table~\ref{Tab:OSG_support}.

\begin{table}
\begin{footnotesize}
\begin{tabular}{|p{1.8cm}|p{9.6cm}|}
 \hline
   {\bf Application Type} & {\bf Characteristics and Examples}
  \\
\hline
Simulation and modeling & CPU-intensive,  large number of  independent jobs, e.g.,
physics Monte Carlo event simulation \\
\hline
Production processing & Significant I/O of data from remote sources and  long sequences of similar jobs passing through
data sets,
e.g., processing of physics raw event data  \\
\hline
Complex workflow & Use of VO-specific higher-level services and
dependencies between tasks, e.g., analysis, text mining \\
\hline
Real time response & Short runs and semi-guaranteed response times, e.g.,
grid operations and monitoring \\
\hline
Small-scale parallelism &
Allocation of multiple CPUs simultaneously \& use of MPI libraries; e.g., protein analysis, MD \\
\hline
\end{tabular} 
\end{footnotesize}
\caption{Types of applications running on Open Science Grid\label{Tab:OSG_apps}}
\end{table}

\begin{table}
\begin{footnotesize}
\begin{tabular}{|p{1.8cm}|p{4.6cm}|p{4.6cm}|}
\hline
& Support & Challenges
\\ \hline
Simulation and Modeling & Batch-system services and prioritization policies;
small amount of data storage & Ensuring full usage of dynamically available resources wherever they are located
\\ \hline
Production processing & Job and workload management tools; data placement and access management tools & \multirow{2}{*}{
\hspace{-0.3cm}\begin{tabular}{p{4.6cm}}
Automation of conditional workflows, retries, etc.; common tools for efficient placement and co-location of data and jobs; support for VO-defined policies applied effectively across the autonomous resources
\end{tabular}
}
\\ \cline{1-2}
Complex workflow & Tools for managing workflows; pre-placement of application tools and databases at remote sites; tools for error reporting, response and tracking &
\\ \hline
Real time response
& Prioritization services to allow immediate or minimum latency execution of jobs & Support for checkpointing and restart of other applications;
dynamic nature of available set of resources precludes deterministic response times
\\ \hline
Small-scale parallelism &
Local support for MPI;  OSG support for publishing necessary information of site-specific configurations and software versions &
Automated use across multiple MPI site configurations and implementations
\\ \hline
\end{tabular}
\end{footnotesize}
\caption{Support for OSG Application Types\label{Tab:OSG_support}}
\end{table}

The OSG provides resource information and matchmaking software for
automated selection of remote sites on which to execute jobs. Users
embed interfaces to this information and/or do manual selection of
sites. Such selections are configured to match the processing and
storage needs and timelines of the applications.

\subsubsection{Successes and Limitations}

OSG has successfully worked
with the US high energy physics (HEP) community and EGEE to build an
infrastructure that allows both HEP and processing for other science and research.  The
challenges
OSG faces include 
meeting the planned (and anticipating the unplanned) capacity and
capability needs of the current user communities; 
managing and accommodating heterogeneity across
facilities that scale from small university department clusters to 
leadership-class computing facilities, with 
user communities that scale from individual PIs and students to very
large collaborations; 
and developing and measuring an agreed-upon, sustainable
economic model for growth that takes account of OSG's bartering and
brokering approach.
OSG best supports loosely coupled applications as well as small
parallel applications that fit on a single multicore CPU.

 \subsection{EGEE and EGI\label{egee}} 

The Enabling Grids for
e-Science \cite{egee} project
supported a multidisciplinary research community that primarily
performs high-throughput
data analysis using a distributed storage and computing
infrastructure built from multiple resource providers operating in
different administrative domains, including supporting the
Worldwide LHC Computing Grid in Europe.
Access to this infrastructure was provided through a software layer
(middleware) that abstracted the distributed resources through a
service-oriented architecture (SOA) into an environment that could be
used as a platform for high-throughput data analysis. The middleware
distribution used within the EGEE project was gLite~\cite{gLite}, an
assembly of software components developed within the project and by
its collaborators.

EGEE is no longer active; it was recently replaced through the European Grid Initiative (a community-driven process with the aim of establishing a sustainable European infrastructure) to provide the European Grid Infrastructure (EGI). The EGI-InSPIRE project has supported the EGI~\cite{egi,egiURL} since May 2010, and during its first year has focused on the transition from a regional to a national operational structure.
This section mainly describes EGEE.

\minihead{History}
EGEE had its origins in the European Data Grid (EDG) project that ran
between 2001 and 2004. EDG's main role was to prototype the
technologies needed to build a European grid infrastructure and to
bring together the groups providing the resources, constituting the
user community, and building the technology components. As
a result of this successful prototyping activity, the EGEE 
projects (EGEE-I and EGEE-II ran between 2004 and 2008) funded by
the European Commission's Framework Programs
were established
to move the experimental grid infrastructure to production
quality. This goal was successfully achieved, and the EGEE-III project
continued the operation of the production infrastructure and preparing
for its transition to a sustainable structure for future production
operation (EGI), while supporting a multidisciplinary community
of 13,000 users across the high energy physics, life sciences,
astronomy, astrophysics, computational chemistry, Earth sciences,
fusion, and computer science domains.
 
\minihead{Mission/Vision}
EGEE's mission was twofold: (1)  to provide a generic production-quality grid infrastructure that was 
continuously available to reliably support multiple user communities and (2) to provide an 
integrated pool of resources to researchers in Europe and their international 
collaborators.  The focus in EGI now is primarily on the operational infrastructure
delivered in collaboration with national grid initiatives and European
intergovernmental research organizations, which are seen as the
main building blocks of long-term sustainability.
 
\minihead{Management}
EGEE's management structures were focused on two issues: the overall
direction and management of the project, which had activities beyond just
running the infrastructure, and the delivery of the
production grid infrastructure itself. EGEE was managed on a daily basis by
the managers of each activity within the project that encompassed the
dissemination, training, user community activities, operations,
networking, software integration, and software development activities
in the project. This approach ensured regular coordination among all the
activities at a managerial level to resolve any technical issues.
 
The delivery of the operational production infrastructure was managed
through regional operational centers (ROCs). ROCs integrated the
resources within a single country (e.g., Italy) or across a large
number of countries (e.g., central Europe or southeast Europe).
Within each ROC, operational teams monitored the state of their
federated resources, identified performance or failed services, and
raised ``trouble tickets'' with the relevant resource providers to
trigger resolution of these problems.

Within EGI, these management structures have evolved to clearly
defined coordination functions established within a new dedicated
organization~\cite{egiURL} that federates an operation infrastructure
contributed by over 35 European national resource providers that
have replaced the regional model established within EGEE.

\minihead{Roadmap/Future}
EGI, a collaboration rooted in the EGEE community and related regional
infrastructure projects such as BalticGrid, SEE-Grid, and the Nordic
DataGrid Facility, is now coordinating the provision of a
European-wide production infrastructure integrated with
production infrastructures around the world as required by its user
community, open to all disciplines. This
moves the support of the infrastructure from a series of short-term
projects to a model that is
more sustainable long-term by leveraging established national and domain-specific infrastructures.

Part of the goal of EGI is to provide greater
integration between high-performance, commodity computing
(grids) and volunteer desktop resources and to include new resources
such as cloud
computing, as increasingly demanded by its users.
Ideally, a single authentication token and interoperable software
distributions (coordinated by the European Middleware Initiative, or EMI)
will eventually provide
secure, controlled, integrated access to all resources regardless of type
and irrespective of
the provider being run by a local, national or international body.
Progress on these two aspects will provide the integrated e-infrastructure
(or cyberinfrastructure) that has been the vision of this community over
the past decade.

The choice of a set of interoperable middleware stacks (gLite, UNICORE,
and ARC) that are supported by EMI and by the Initiative for Globus in
Europe (IGE), rather than a single, monolithic distribution, was made
because different user communities (including new communities that EGI
wants to attract) have different needs that can best be met through
different technologies.
Additionally, while some solutions are comparable, they may be adopted
by different sites or countries for nontechnical reasons. For
sustainability, most of the larger EGI sites will likely end up having
to support multiple communities and so will have to support multiple
stacks. The integration and harmonization activities being undertaken
within the EMI project may reduce the number of stacks that
eventually need to be deployed.
 
\subsubsection{Characteristics}

EGEE supported a user community that ran applications
from research domains as
diverse as multimedia, finance, archaeology, and civil protection. The
users benefited from a federated distributed computing infrastructure that
operated around the clock across approximately 300 sites in 50 countries,
encompassing 140,000 CPU cores and many petabytes of on- and near-line
storage.
 
The applications run on EGEE focused on the computational analysis and generation
of data stored within the EGEE infrastructure. In some cases, this data was stored
remotely from where the analysis was performed; mechanisms were
provided to move the data or place the computational analysis near to
the data location. In other cases, data was replicated throughout the grid,
allowing jobs to retrieve data from or locate themselves near a
particular copy. While many of the applications were executed on a single
core, support was also provided for parallel applications (MPI) on
resources that were enabled to support this workload.

The following were key aspects of EGEE:
\begin{shortlist}
\item Exposing the grid resources: computing and storage elements hosted by
the resource providers that were part of EGEE advertised their resources
in the information index. 
\item Controlled access: not every community or project, represented
  by one or more virtual organizations, had access to every resource
  within the grid. An individual's role in a virtual organization
  was managed through a service (VOMS) that specified the roles a user
  had within that organization.
\item Consistent availability: the grid fabric was monitored to ensure its
availability through tests that were, in addition, able to determine the
version of the installed software.
\end{shortlist}

\subsubsection{Usage Modes}

The key function of EGEE was to manage data files located
on storage elements throughout the grid. Data files could be
registered in a file catalogue where their physical location could be
mapped from a logical name. Multiple physical
copies of a data file could be distributed within the grid, mapped from a
single logical name. Physical data files could be moved between
storage elements, which could encompass temporary or permanent disk
caches or near-line tape storage, as part of the data
analysis. Applications were then deployed on EGEE and used to
analyze the data. 
 
The  EGEE Grid infrastructure (using the gLite middleware) was
developed to support high-throughput computing, where the work could be
based around the movement of data (files) as part of a computational
analysis workflow. Work could be submitted directly to a computational
element by a user or through the Workload Management Service (WMS)
that selected a resource according to the requirements of the
application specified by the user and the then available resources
that the user had access to through virtual organization
membership(s). Physical copies of a logical data file could be located
through a file catalogue. The movement of files was coordinated
through a file transfer service that enabled policy to be imposed on
the use of dedicated network paths linking the transfer sites.
 
\subsubsection{Successes and Limitations}
 
EGEE provided a production-quality infrastructure to its community. It
supported the four experiments using the Large Hadron Collider,
the life sciences community through medical imaging, bioinformatics
and drug discovery,\footnote{See, for example, the recent WISDOM
experiments \url{http://www.isgtw.org/?pid=1000993}.} and many other
application communities.\footnote{See ISGTW for other examples:
\url{http://www.isgtw.org/}.}  EGEE collaborated with OSG to provide
interoperating federated infrastructures that could be used transparently
by the LHC experiments' software.
EGEE found limitations in scalability,
reliability, and efficiency, which it worked to overcome during its
seven-year, multinational development effort.

 \subsection{The UK National Grid Service}

The UK National Grid Service (NGS) is a national consortium of computational and data resources that use defined, open standard grid interfaces to provide services to academia. The NGS Collaboration is made up of the universities of Belfast, Birmingham, Bristol, Brunel, Cardiff, Durham, Edinburgh, Glasgow, Imperial College, Keele, Lancaster, Leeds, Liverpool, Manchester, Oxford, Reading, Royal Holloway (University of London), Sheffield, Southampton, Westminster, and York; Rutherford Appleton \& Daresbury Laboratories (STFC); HPCx and HeCTOR (supercomputers); and the Welcome Trust Sanger Centre. 

\minihead{History}
The NGS has been operational for over four years, providing the underpinnings of a national e-infrastructure (for the UK) through the establishment of the solid management methods and interface definitions necessary to allow users to access the available resources.

The first phase (2004--2006) funded four sites that each hosted resources, split into two ``compute'' systems and two ``data'' systems. These were complemented by a support center that administered systems for managing user access, information aggregation, and monitoring, as well as providing end-user support and training. The basic technical coordination and management structure of the project was also defined at this point. The core resources were upgraded during phase two (starting 2006) of the project, with identical compute systems at each site, in addition to the existing data systems.

\minihead{Mission/Vision}
The UK NGS mission statement is ``to enable coherent electronic access for {\em all} UK researchers to {\em all} computational and data based resources and facilities required to carry out their research, independent of resource or researcher location.'' The goals are as follows:

\begin{shortlist}
\item Enable a production-quality e-infrastructure.
\item Deliver core services and support. 
\item Integrate with international infrastructures following user community demand.
\end{shortlist}

This final goal links to the NGS becoming the UK representative within the European distributed e-infrastructure project, EGI.

\minihead{Management}
Following the successful commissioning of the core resources, the NGS has expanded significantly through contributions from partner institutions. This expansion has led to the development of policies and procedures to ensure the consistent quality of resources that are attached to the NGS.
A resource can join the NGS at two levels: partner or affiliate. A partner should provide a `significant resource or service' for NGS users. The procedure for joining is defined such that after notifying the NGS of their intention to join, sites gain assistance in installing the necessary software interfaces. Once installed, these must complete a full week of compliance testing without error before being certified as a recognized NGS resource. Additionally, they must complete a service-level description (SLD), detailing the resource and level of support they intend to offer users. They are also eligible to nominate a representative from their organization to attend the NGS Technical Board.

In contrast, an affiliate, while still having to pass interface and service tests, does not have to provide an SLD and may maintain control over the user community that is being  served. As of November 1, 2010, 30 institutions were members of the NGS, with 10 partners and 18 affiliates including the national HPC resources and 7 institutions that are community members.

\minihead{Roadmap/Future}
The UK NGS has been nominated by its funding agency as the UK representative within the European EGI project. The result has been a closer integration between the NGS and the GridPP\footnote{GridPP is a collaboration of particle physicists and computer scientists from the UK and CERN who have built a distributed computing grid across the UK for particle physicists.} project, which has until recently been the main method of UK engagement with European grids. (GridPP is an example of a community self-organizing to provide resources that their users need; and as more large research infrastructures are built throughout Europe to which UK researchers will need access, other communities may follow suit.) Through the alignment of core functions, shared services are provided to the two grids. The NGS will continue to provide the central services that are shared by communities and, as such, are aligning with services required for performing its nominated EGI central functions, as well as national versions of other services.

\subsubsection{Characteristics}

The NGS provides a single-point-of-contact HelpDesk for support and queries, for example, digital certificate issues and requests for new application software.

The NGS has a set of interfaces defined through a ``core software stack'' developed with a desire to maintain compatibility with other large infrastructures such as GridPP and EGEE. This has meant defining an interface for which a number of software solutions can be used. The solution chosen by the core nodes has included the usage of the pre-Web Services version of the Globus Toolkit~\cite{GT} (GT 2) packaged within the Virtual Data Toolkit~\cite{VDT}. The interfaces provided include job submission, information publishing, data movement, and grid security infrastructure-based (GSI-based) secure shell access. This is one of a number of solutions that the NGS has documented and that are available for a site to install. Other middleware that can be installed before obtaining NGS membership includes GridSAM~\cite{gridsam} and Globus Toolkit version 4 (GT 4, with Web Services.)

Each NGS installation is tested at regular intervals to ensure compliance, using an INCA-based~\cite{INCA} monitoring framework. Building on the lower-level services provided by the middleware, the NGS also has a number of different services that provide higher-level functionality. These include resource brokering, preconfigured application portals, and resource information publishing. Overall, the NGS has a managed approach to change, providing stable, robust services and supporting them over a reasonable period. At the same time, it is recognized that new services need to be developed, deployed, and supported for the future growth of the NGS; therefore, the communities can depend on the services NGS provides in the longer term.

Although paid-for, or subscription, services are possible, current NGS services are free at the point of use, funded by the UK funding agencies EPSRC, JISC, and CCLRC (now STFC). 

As of late 2010, the NGS had about 1,200 users, about 80 of whom submit jobs in any given month, and with significant usage in computer science, chemistry, physics, biology, engineering, biochemistry, informatics, mathematics, and medicine.  In 2010, the NGS supplied about 600 CPU-days of computing per day, about one fourth of which were submitted through Globus~\cite{ngs-data}. 

\subsubsection{Usage Modes}

The NGS user communities operate on the system in a number of different modes depending on the type of resources they use and the type of underlying problem they are working on. Those using the resources primarily for computing can submit a task to a resource, either a prechosen system or one automatically selected through a resource-matching or brokering functionality. These jobs may be one of a number of independent tasks or a single parallel job that is run on a single ``MPI or OpenMP''-capable resource.

A number of different types of portal systems also are available. The first of these is the NGS Application Repository, which makes a number of preconfigured applications available through a JSR 168 compliant portal framework. Here, the NGS installs the applications and sets up the appropriate pages within the portal for the application. An alternative is the Application Hosting Environment (AHE)~\cite{ahe}, which assumes that a research group or community has an expert who is able to configure and install the applications needed and will then make them available to the rest of the group. AHE can submit jobs to a number of the resources on the NGS and is currently being used by groups in biochemistry, chemistry, and materials science. NGS users also use other systems for automating job submission and management, including the parameter sweep managing Nimrod/G system~\cite{nimrodG} and the GridBS~\cite{gridBS} resource broker. These are intended primarily for users or institutional communities to submit tasks to the full range of resources to which they have access. The NGS is also deploying further high-level services, such as programming abstractions, that users can use in their software systems. In addition to using grid-type interfaces for job submission, a significant number of users access NGS resources by logging in to the end system using a single sign-on-enabled version of SSH and interact directly with the local distributed resource manager. These users often come from institutions with overloaded HPC resources, HPC resources with charges for usage, or no HPC system. 

The applications that run on the NGS are wide and diverse. Within the HPC communities, a significant number of users are using commercial or community codes. Within the HTC community of users, the situation is almost completely reversed, with the majority using their own developed codes, though these may depend on commercial or community libraries. Thus, these applications are more easily distributed around the grid systems, particularly when they are statically linked so that version interoperability difficulties for libraries are minimized. 

The communities that have used the NGS have been extremely broad, from STEM (Science, Technology, Engineering, and Mathematics) to art, humanities, and social sciences. Well-known examples have been used to create case studies to publicize the ongoing user communities. This approach has been particularly effective because of researchers being much more willing to listen to ``their own'' than to a set of service operators or even their own institutional computing services.

Recent work has also included the provisioning of a test system for cloud-type services, with the intention that using this technology will allow for user services such as clients, portals, and workflow engines.  These will be installed, demonstrated, and used by communities who wish to have a unified software face to their collaborations and work but who feel that installations on desktops and local resources are too difficult or time consuming. They also may have licensing restrictions that limit how useful the software would be for a whole community.

\subsubsection{Successes and Limitations}

The NGS has attracted about 700 users from a wide variety of academic fields (e.g., biology, physics, and computing) with a variety of computational and data problems (e.g., simulation of UK population dynamics) and ranging from part of a large collaboration to the individual researcher. The enabling effect of NGS resources has been acknowledged in a significant number of academic publications. 

Overall, the NGS has been extremely successful, although because of the way that the UK has developed two parallel grid systems, user communities have sometimes been confused. The GridPP system is a significant contributor to the EGEE system and, as such, has a large community of users. There are also a significant number of EC-funded projects that should be making use of e-infrastructure but possibly because of the duality, the UK contributors are not making use of the NGS. Also, a significant investment in the UK university sector in mid-range HPC systems has led to a number of NGS users moving to these systems.
Overall, it appears while a significant number of users want to use HTC and grid-type computing, they often need significantly more user support than the NGS is funded to provide. To counter this situation, the NGS is engaging community and institutional champions to enable communities to support themselves.  

Additionally, licensing can be a significant impediment to the use of some applications on some NGS resources. Users can work around licensing issues, however,  by binary building, distribution of runtime environments, and use of open source compatible equivalents.

\section{Research Production Distributed Infrastructures\label{sec:rdis}}

In this section we discuss four national and international research production distributed infrastructures.

 \subsection{Grid'5000}

Grid'5000~\cite{grid5000} has been designed
as a highly reconfigurable experimental testbed for
large-scale distributed systems.  It includes more than 5,000 cores
in clusters at nine sites
across France, connected by a network with dedicated capacity.

\minihead{History}
Preparation for Grid'5000 began in 2003 with a series of interviews
of 10 research groups active in grid computing in France.  These
10 groups described 100 potential experiments.  In general, the experiments
were diverse in their infrastructure needs, a situation that was reflected in the
design of the infrastructure, which entered production in 2005.

\minihead{Mission/Vision}
Grid'5000 is designed to support
experiment-driven research in all areas of computer science related to
parallel, large-scale, or distributed computing and networking.
Experiments that use Grid'5000 should lead to results in those research
fields and use the resources as a model for the use of nonacademic resources.
Available resources can be used in a low-priority mode to generate
useful results for other communities, especially if this generates
results that are also relevant to the main research fields of
Grid'5000.

The initial Grid'5000 machines are distributed across nine 
sites, a side effect of the way the construction of the
Grid'5000 was funded.  Because securing resources for large-scale
experiments (at least three sites and 1,000 CPUs) can be difficult in the
absence of specific rules and because these experiments are a driving
factor for a multisite instrument, such experiments are favored by
Grid'5000. Nevertheless, research at a smaller (local) scale is also
welcome.

\minihead{Management}
The Grid'5000 executive committee 
(the scientific director, the deputy scientific director, the
technical director, representatives from each Grid'5000 site, and a representative
of RENATER, the French National Research and Education
Network provider) meets once a
month by teleconference.  Directions for the technical team's work are
laid out in a document written in 2008 for the next four years under the
technical director's leadership and reviewed by the executive
committee. This document allocates resources to the technical team,
and an updated workplan is submitted every year using the same
process.  A steering committee, representing the funding institutions,
meets once a year to review the board of directors' action and to give
recommendations on the directions to take.

\minihead{Roadmap/Future}
Grid'5000 has become a tool for everyday work for the
research community in France,
and it has been classified
a very large research infrastructure by the French Ministry of
Research. The institutional context of Grid'5000 is evolving
to ensure the sustainability of Grid'5000 and especially the renewal of
the hardware used to run some sites.
Specifically, three major activities are under way. The first is work on
the network links between sites, to enable bandwidth reservation and
measurement at a fine-grained level. The second is extending Grid'5000 to
new sites. A memorandum of understanding has been signed with Porto
Alegre, Brazil, and additional sites are in preparation. The
third is development of an API to improve the scriptability of working on
Grid'5000. 

\subsubsection{Characteristics}

Grid'5000 comprises a number of sites interconnected by a
dedicated network.  A Grid'5000 site has two attributes:
(1) a single LAN with a frontend and (optionally) an access machine, a
server for the resource scheduler and a server for deployment, an NFS
server and a DNS server, and a route to the interconnect network and (2)
One or more clusters of machines. The objective is to have at least
128 nodes per site.

A site manages Grid'5000 machines and possibly other resources. These
other resources are considered to be outside 
Grid'5000 
but are integrated in the site (with the same accounts and same
resource scheduler) because they are useful to the community.  The
resources of a site are static and are described in the resources
scheduler's database.  Thus, volatile sites are excluded from Grid'5000;
sites are either available or going through maintenance operations.
Requiring that resources on a site be static avoids having to manage
dynamic addition and retrieval of resources and limits the
complexity of the testbed for users.  For sites that want to put
resource sharing with other projects in place, periods where the
resources are made available to other projects are required to appear
as reservations of the resources, and their existence must be
negotiated with Grid'5000.

A Grid'5000 system has the following properties: it is exclusively
available in Grid'5000 context; it can be allocated to users without
requiring the use of specific properties during job submission; and it is
managed by kadeploy~\cite{kadeploy} in a reliable way---that is, it can be managed
remotely (reboot, power-off, power-on, etc.).

Some systems can have unusual properties, and therefore the resource
scheduler can be configured so that these systems are last to be generally
allocated. Moreover, some users can be given higher priority to access
these specific systems if required for day-to-day work.

Accounts are requested by users at one of the sites participating in
Grid'5000 and are approved by the site's chief scientist. This approach gives
users complete access to all the resources of Grid'5000 at all sites
without any usage quotas, as well as disk space on the NFS server
serving home directories on each site.

A tool tracks usage and relates it to reports that users have to
update regularly. The reports describe planned usage, current usage,
and results obtained using Grid'5000 
and are published on the website.

In 2009, Grid'5000 was used by 572 different people, with an average
of 272 different users over a three-month period. Of these 272 unique
users, one-third used three or more sites on the same
day.

{\it Network}: Because reproducibility of experiments is a goal for
Grid'5000, the network interconnect 
is dedicated, ensuring the only perturbations seen in the interconnect
links are those generated by the testbed. Because experimenting with
the network layer of large-scale distributed systems, including
testing new protocols, is a goal for the testbed, the interconnection
provides a layer 2 service.  The first generation of interconnect used
Ethernet over MPLS-VPNs between all sites. It was a full mesh topology
based on MPLS tunnels established between the RENATER POPS and
the Grid'5000 sites. In practice, sites were interconnected through
1-Gbps VLANs.  The current version uses a dark fiber infrastructure
allowing for 10-Gbps links. With this infrastructure, Grid'5000 sites
are directly connected to switches inside RENATER POPs and see each
other inside the same VLAN.

{\it Site independence}: Grid'5000 systems do not have special
provisions to guarantee high availability.  Demands for electricity,
network, and cooling equipments are such that machines at any one site
can remain unavailable or unconnected to the others for a few
consecutive days every year because of maintenance or upgrade
operations.  Such operations should have only minimal impact on the
availability of resources. Of particular importance are the following:

\begin{enumerate}
\item Machines hosted on a site that has no Grid'5000 network
  connection to the other sites should still be usable by all users
  who have access to the site using an out-of-band network connection.
  This particularly concerns users from the site hosting the machines.

\item If any site has no Grid'5000 network connection, the other
  Grid'5000 resources should remain usable, even to users at that site
  who have access to other sites using an out-of-band network
  connection, through public access points for example.
\end{enumerate}

This design decision has proven valuable for the day-to-day
availability of Grid'5000 resources. It also has profound impact on
account and resource management.

For account management, this design decision implies the following:

\begin{itemize}
\item A distributed architecture is needed for
  authentication and authorization. A master LDAP server holds all account
  information and is replicated on a slave server on each site. This
  slave server remains functional even after having lost its
  connection to the master server for a few days.
\item A different home directory must exist on each site for a given
  user. No automatic synchronization is provided to users, but one of
  the first tutorials explains to users how they can synchronize their
  data.
\end{itemize}

For resource management, this design decision implies the use of a
independent resource scheduler on each site. This, in turn, leads to
co-allocation problems when users try to run experiments spanning
multiple sites. In order to handle this problem, advance reservation of
resources is required from the resource scheduler.

{\it Reconfiguration}: At the core of Grid'5000 concepts is
reconfiguration of resources by users. The motivation is
to have an instrument on which all existing Grid middleware can be
deployed by users and therefore compared.  In the current iteration
of the infrastructure, only reconfiguration of the software stack of
nodes is possible. Users have to choose one node on each site they use
to act as a head node for their experiment, if applicable, to mimic a
classical grid environment.

The concept is to give users complete control of Grid'5000 nodes by
allowing them to deploy their complete environment, including the
operating system or hypervisor, on nodes they are allocated. This is
done by changing the contents of hard drive on these nodes, as well as
the nodes' PXE directives, used to boot using kadeploy.

As user-deployed environments are, by definition, not controlled, the
reconfiguration tool cannot make assumptions about the deployed
operating system, and might not even be able to log into the
environment at the end of a job to restore the node to a default
state. Therefore, reconfiguration requires hardware support, in the
form of management cards on the nodes.  Grid'5000 has found that these
management cards need their own independent access to the
network. 

Grid'5000 provides either a default environment or  a seed environment to users that boots on
all Grid'5000 hardware.
Users can
customize the seed environment according to their needs.

\subsubsection{Usage Modes}

Grid'5000 was built to be used for a wide variety of experiments on
large-scale distributed systems.  (An experiment is typically composed
of one or more jobs running on Grid'5000's clusters.)  One of the key
issues is how experiments are prepared and run.

In Grid'5000, users develop and debug their experiments during normal
work hours, with all the resources available for large-scale
experiments during the night and during weekends. These resources can be viewed
as a network of workstations, but the other way round: all machines
are part of a local cluster but are made available to users during
work hours. This approach helps build an infrastructure with a very large
number of nodes.  
It should be understood as applying locally to each cluster, therefore
shortening the time all resources can be used at the same time for a
single experiment.

The target life-cycle of an experiment involves three phases: (1) develop the
software stack to experiment on a few machines on one or more sites,
in interactive mode during normal work hours; (2) automate running of the
experiment by developing ad hoc scripts; and (3) run the experiment in
batch mode using an increasingly large number of resources, outside
normal work hours.  This  has proven difficult to promote efficiently,
however,
because users tend to skip the second step as the resource scheduler
implements interactive and advance reservations. Users tend to simply
reserve in advance an increasingly large number of nodes interactively
to run their experiments, preferring to stay long hours rather than scripting
the experiments. 

Grid'5000 has two modes: submission, in which the user submits an
experiment and lets the scheduler decide when to run it, and
reservations, in which the user makes a reservation to run an experiment at a
specific time and then, at that time, launches the experiment
interactively.

A second class of experiments has also emerged: CPU-hungry users who
are eager to fill in any gap in the scheduling of resources to
run a specific experiment. Those users are allowed use of resources
in best-effort mode, where their job will be killed if anybody
requests the resources. No infrastructure has been built to cater to
their specific needs, and this situation could be problematic for fair sharing
between these experiments. For the time being, this is handled in an
ad hoc fashion, where users require approval of the experiment in
advance to discuss the way this sharing will be implemented.

{\it Principles}: Grid'5000 is a shared system, used by many people
with different needs. The administrators pursue two
objectives.  First and most important, they want to make Grid'5000 available
to experiments involving a significant number of nodes (in the
thousands). In order to make this possible, reservation fragmentation must be
avoided as much as possible. Second, they seek to keep Grid'5000 available
during the day for the development of experiments. Therefore,
reservations using all the nodes available on one site during work
hours for that site should generally be avoided. 

\subsubsection{Successes and Failures}
More than 600 experiments have been executed on the platform since it
was made available to the community. These have led to more than 400
publications in international journals and conferences and over 30 PhD
theses defended. Many experiments used more than five sites and more
than 1,000 nodes. From low-level network protocols to ``classical''
application parallelization and grid middleware large-scale
validation, Grid'5000 has become a
highly valued target evaluation platform for computer science. Some records were broken
with Grid'5000. For example, the prime factors of the RSA challenge
number RSA-768, using the Number Field Sieve, were obtained by an
international team of scientists from EPFL (Switzerland), INRIA
(France), NTT (Japan), CWI (the Netherlands), and Bonn University
(Germany). The calculation took less than 2,000 core-years on modern
CPUs (including the nodes from the Grid'5000 platform).

{\it Diversity of sites and cultures}: As many grid projects have
found out, one of the most difficult tasks when building a distributed
architecture is to have local cultures converge. This is especially
true for system administrators.  Ideally, all system administrators
should be able to help out to manage distant sites. But this approach can 
be efficient only if all sites share a common architecture and server
distribution, which is not possible because each site depends
on an independent
administration and local funding.
This in turn clashes with local strategies, where
Grid'5000 site administrators also share their time between
administration of other machines of their lab.

Efficient support, in terms of manpower use or of quality of service,
for a distributed testbed remains an open issue.  In the first
Grid'5000 phase, every site had to find local manpower to manage the
Grid'5000 hosted locally. 
All system administrators had to install and configure every needed
service for their own site and often applied local strategies for
administration. It could therefore take time for an update to be
applied on all sites, thus reducing the coherence of 
Grid'5000 as experienced by its users. Moreover, this organization
encourages system administrators to think locally.

In the second phase, a dedicated system administration team of five
people was created with access to eight of the nine sites. This eases
the quick deployment of updates as well as a ``think global'' attitude to
system administration. Nevertheless, physical access to the machines
is frequently needed; and for sites with no member of the team
present, complex interactions with local staff are necessary.  This
second phase has increased automation of tasks and could lead to a
third phase where system administration task are automated using a
central configuration management tool. One could then imagine part-time
system administrators on all sites and a core team to manage all
sites. The drawback to this strategy is that it could kill local
knowledge of cluster and experimental machine administration on sites
not hosting the core team.

{\it About usage patterns}: Because one of the aims of Grid'5000 is large
scale experiments, resource fragmentation in many small experiments is a major
concern that is exacerbated by long-running
jobs.  As people attempt to share the cost of clusters
between Grid'5000 and other projects, incompatible usage patterns are
often an issue. The policy of Grid'5000 is to have frequent but short
periods of time where all the resources can be given to an unique
user. This can prevent the effective sharing of resources with users
who want to run jobs that last a week or more.  Understanding the
expected usage pattern of the target users and setting up rules to
enforce them have proved crucial for the success of Grid'5000.

 \subsection{PlanetLab}

PlanetLab ~\cite{planetlab} is an open, globally distributed platform for developing, deploying, and accessing planetary-scale network services. It has been used primarily as a research and education testbed for distributed computing services and applications.

\minihead{History}
In March 2002, a small community of researchers interested in
planetary-scale network services proposed PlanetLab as a community testbed. The initial 
participants were Berkeley, MIT, Washington, Rice, Princeton, Columbia, Duke, Carnegie Mellon, and Utah. 
Intel Research provided the initial 100 machines, which by October 2002 spanned 42 sites.
In February 2003, PlanetLab nodes came online at 
three of the points of presence (PoPs, or access points for the network) on Internet2's Abilene backbone. All 11 Abilene PoPs were hosting PlanetLab nodes by the 
end of 2003. In 2003, NSF announced a \$4.5M award to Princeton, UC Berkeley, and Washington 
for supporting and enhancing PlanetLab.
 In January 2004, Princeton, Berkeley, and Washington formally created the PlanetLab Consortium, 
with Intel and HP as charter commercial members. Princeton began hosting the consortium, and 
operational responsibility for PlanetLab moved from Intel to Princeton. By June 2007, 
PlanetLab passed the 800-node mark.  In July 2007, PlanetLab federated with the OneLab 
project, which began to support PlanetLab-Europe (PlanetLab-EU). As of mid-2010, PlanetLab had 1,132 nodes at 518 sites.

\minihead{Mission/Vision}
PlanetLab's goal is to support both experiments (short-term)
and network services (continuously running)
and ultimately to develop and demonstrate a new set 
of network services at planetary scale. 

\minihead{Management}
The PlanetLab Consortium is a collection of academic, industrial, and government institutions 
cooperating to support and enhance the PlanetLab overlay network. It is responsible for 
overseeing the long-term growth of PlanetLab's hardware infrastructure, designing and evolving 
its software architecture, providing day-to-day operational support, and defining policies 
that govern appropriate use. Institutions join the consortium by signing a membership 
agreement and connecting two or more nodes to the PlanetLab infrastructure. 
A governance document describes how the consortium is organized. 

\minihead{Roadmap/Future}
PlanetLab is in the early stages of federation; it is creating autonomous authorities 
that are responsible for subsets of the global slices and nodes. These authorities will then 
peer with each other to build a federated system. This effort is being done with an eye 
to eventual federation across a collection of testbeds. One of these autonomous authorities, the OneLab 
project, will create independent slice and management authorities spanning Europe.
Universit\'{e} Pierre et Marie Curie (UPMC) will run a subauthority
(PlanetLab-EU) under the PlanetLab root authority. PlanetLab-EU is expected to
operate in a way that is consistent with the PlanetLab's primary mission as a global
testbed for developing, deploying, and accessing planetary-scale network services, but it
will otherwise be an independent management authority (responsible for the stability of
a set of nodes) and slice authority (responsible for the behavior of a set of slices).
PlanetLab and PlanetLab-EU will run independent operations
teams, although the two teams will
work to define a common response procedure and template messages.

\subsubsection{Characteristics}

PlanetLab is a collection of machines distributed over the globe.
Most of the machines are hosted by research institutions, although some are in colocation and routing centers (e.g., on Internet2's Abilene backbone).

PlanetLab has a common software package. 
All PlanetLab machines run this package, which includes a Linux-based operating 
system; mechanisms for bootstrapping nodes and distributing software updates; a collection of 
management tools that monitor node health, audit system activity, and control system 
parameters; and a facility for managing user accounts and distributing keys.

PlanetLab supports running short-term experiments, 
as well as long-running services that support a client base.

PlanetLab is a 
microcosm of the next Internet. Not only are researchers evaluating and deploying end-user 
services on top of PlanetLab, but they are also expected to develop foundational subservices 
that can be folded back into PlanetLab, thereby enhancing the facility for others.

Researchers who make claims about protocols and services running on the Internet use 
PlanetLab to demonstrate how their designs hold up under realistic network conditions. 

PlanetLab has hosted over 4,700 users in its six-year history, approximately 3,700 of whom have 
been students. Whether these students are working on their PhD research or doing course 
assignments, they are gaining valuable experience with network systems running at a global 
scale---including coping with transient failures, differences in connectivity cliques, 
variations in latency and bandwidth, and abuses (some of which are malicious)
inflicted by real users.

Additionally, a set of graduate and undergraduate courses have been 
designed to take advantage of PlanetLab.

\subsubsection{Usage Modes}

PlanetLab is used primarily by systems researchers to understand the
requirements for deploying network services (e.g., resource discovery,
network protocols, content distribution, P2P routing). Many
concurrent experiments are run across the shared infrastructure. PlanetLab applications typically exploit the wide-area connectivity provided by its many sites, for example, new peer-to-peer systems and applications. A user acquires a slice (a set of nodes), deploys onto the slice, then releases it when the experiment is done.  Some applications run on a small slice to test small-scale services (tens to hundreds of nodes). Other applications, such as monitoring and content distribution, tend to run on all of the available nodes providing the largest degree of network coverage. These large applications tend to be long-running or persistent, while the smaller-scale services are generally used for short-term experiments and are transient.

\subsubsection{Successes and Limitations}
PlanetLab has proven to be a valuable platform for learning about network-wide phenomena, 
creating new network protocols, evaluating new and existing network services, gaining 
experience with network systems running at a global scale, and deploying novel 
network services that enhance the capabilities of the Internet.  PlanetLab also has formed the basis for NSF's GENI
initiative~\cite{GENI} into new Internet designs.
Quantifying the broader impact of this research is difficult, but anecdotal 
evidence strongly indicates that research leveraging PlanetLab is having a far-reaching impact. The following is a s small sample. 
\begin{shortlist}
\item The iPlane~\cite{iPlane} and Hubble network measurements~\cite{hubble} projects have been a valuable resource 
for the networking research community, with more than 20 research projects using the 
structured network topology information produced by the systems.  
\item BitTyrant~\cite{BitTyrant} is a highly optimized and strategic BitTorrent client whose development 
was aided by extensive experimentation on PlanetLab.  BitTyrant was publicly released in 2007 and was downloaded by more than a million users in its first year.
\item PlanetLab was used for experimentation with localizing optimizations for
peer-to-peer systems.  Out of this work came a new proposal,
P4P~\cite{P4P}, an interface that allows ISPs and 
peer-to-peer systems to coordinate and optimize for both network-level efficiency and 
application-level performance.  
\end{shortlist}

In PlanetLab, however, it is difficult to run repeatable
experiments because of the lack of resource guarantees. PlanetLab
itself can be volatile, with machine availability fluctuating
wildly. Moreover, little attempt has been made to maintain the health or uptime of
PlanetLab nodes: these activities are left to the sites
themselves. Thus, it can be difficult to get a global picture of the
state of PlanetLab.

 \subsection{DAS\label{Sec:DAS}}

The Distributed ASCI Supercomputer (DAS) is a
Dutch distributed-computing platform aimed at computer science research.

\minihead{History}
The Dutch research school ASCI (Advanced School for Computing and  
Imaging)
has set up four generations of the DAS
system over the past 14 years. Each incarnation consisted of four to six clusters
located at different universities, integrated into a single system.
The systems have been used for over 60 PhD theses and for
numerous large collaborations,
including the 30-40M EURO knowledge infrastructure
projects VL-e~\cite{VL-e} and MultimediaN~\cite{MultimediaN}
and dozens of large national and European projects.
The computer science research done using DAS has shifted focus over  
time, from
cluster computing in DAS-1 starting in 1997, to distributed computing
in DAS-2 starting in 2002, to grid computing and e-Science in DAS-3  
starting in 2006, to hardware diversity and green IT in DAS-4 starting
in 2010.

\minihead{Mission/Vision}
The purpose of DAS is to allow computer science experiments, for  
example,
distributed experiments that use multiple clusters simultaneously;
experiments that need high-speed optical networks;
and experiments for which accurate, reproducible performance  
measurements are required.

\minihead{Management}
The DAS project is managed by a steering committee with staff members
from all participating sites. The committee is in charge of making 
overall decisions about the infrastructure. In addition, ASCI 
has set up a team of highly skilled people (mostly scientific programmers)
from all sites who are in charge of systems management.
An attempt is made to simplify systems management as much as possible,
which has proven to be a successful strategy, resulting in a stable
and reliable environment.

\minihead{Roadmap/Future}
The most recent system, DAS-4, has been operational since October 2010
and will allow experiments
with various types of accelerators such as
GPUs, FPGAs,
multiprocessor system-on-chip (MP-SoC), and many-core processors.
DAS-4 consists of six largely homogeneous clusters extended with
a variety of such accelerators. 
ASTRON (Netherlands Institute for Radio Astronomy) is a new partner
in DAS-4 and brings in data-intensive astronomy applications.

\subsubsection{Characteristics}

DAS differs from production systems in many aspects.
Foremost, it is designed to allow clean, laboratory-like experiments,
as opposed to running large production jobs. The system therefore is
largely homogeneous and uses the same processor type and operating  
system
on all nodes. Also, nearly all clusters have the same local network  
(Myrinet in DAS-1 to DAS-3, InfiniBand in DAS-4).
This simple design results in a reliable, easy-to-maintain system with
reasonably reproducible performance.

DAS is designed to allow distributed experiments that run on multiple  
clusters
at the same time. Therefore, the load of the clusters is deliberately
kept low: only short-running jobs (less than 15 minutes) are allowed  
during
daytime. The usefulness of the computer science research that
can be done with the system is optimized; utilization degree
is not maximized (and to some extent is
even ``minimized'').

DAS-3 and DAS-4 have an optical private network interconnect called StarPlane,
provided by SURFnet,
linking the different sites with multiple, dedicated 10-Gbps light paths.
An important goal of DAS-3 was to investigate how the topology of
such an optical network can be changed dynamically.
A photonic switch is being designed for DAS-4 that will allow
topology changes within seconds.

\subsubsection{Usage Modes}

Over the years,  three broad categories of patterns of usage
that scale well on DAS have been identified:

\begin{shortlist}

\item \textit{Master-worker} or \textit{divide-and-conquer} patterns  
scale well because
they generally have good locality and thus relatively little wide-area
communication.

Examples that have been investigated include medical image analysis,
N-body simulations,
SAT-solvers, gene sequence analysis, and automatic grammar learning.

\item Applications with \textit{asynchronous high-throughput  
communication} perform well
because they can do latency-hiding on the wide-area networks. The  
bandwidth of
the wide-area network usually is less of a problem (especially given
our optical interconnect).

Examples include distributed model checking and search applications.
Many measurements have been done with the DiVinE model checking system~\cite{divine}
on wide-area DAS-3, demonstrating that much larger models can
be validated on a grid than on a single cluster.
Also, the Awari solver~\cite{awari} has been implemented
on wide-area DAS-3~\cite{DAS3}.

\item Applications with mixed task parallelism and data parallelism often also  
scale well
because they can use (often fine-grained) data parallelism within a  
cluster
and (more coarse-grained) task parallelism between clusters.
The best DAS example is multimedia content analysis, with which
many (award-winning) large-scale grid experiments have been done.

\end{shortlist}

DAS has developed its own programming systems, including Ibis, Satin,  
JavaGAT,
SmartSockets, and KOALA:

\begin{shortlist}

\item Ibis~\cite{ibis2010} aims to dramatically simplify
the programming and deployment process of high-performance grid
applications. Its philosophy (``grids as promised'')
is that grid applications should
be developed on a local workstation and simply be launched from
there on hostile grid environments that are dynamic and
heterogeneous and suffer from connectivity problems.
As an example, the CCGrid'08 Scalable Computing Challenge was won
using Ibis to create ``scalable wall-socket multimedia grid computing.''

\item Satin is a programming system based (like Cilk~\cite{cilk}) on divide-and-conquer
parallelism, which transparently handles resource failures and  
malleability.

\item The Java Grid Application Toolkit (JavaGAT)~\cite{javagat} offers a set of  
coordinated,
generic, and flexible APIs for accessing grid services from application
codes, portals, data management systems, and so on. JavaGAT sits between
grid applications and numerous types of grid middleware.

\item The SmartSockets communication library~\cite{smartsockets} automatically
discovers connectivity problems (due to firewalls, network address  
translation,
nonrouted networks, multihoming)
and solves them with as little support from the user as possible.

\item  KOALA~\cite{koala} is a grid scheduler that supports co-allocation of  
multiple clusters
at the same time.
Most DAS applications may run on multiple clusters simultaneously, over a  
short period.
They need an efficient scheduler and support for I/O to stage the
input and result files in and out.

\end{shortlist}

\subsubsection{Successes and Limitations}

Several applications were described above.
In addition, DAS-3 has been used for collaborations between computer  
scientists
and application scientists, for example:

\begin{shortlist}

\item DAS-3 was used to analyze the computational characteristics of the
multiphysics simulations published in \emph{Nature}~\cite{DASsupernova}.
It was discovered that the brightest supernova ever recorded,
SN2006gy, was the result of emergent behavior in a dense star cluster.

 \item The MultimediaN project has used DAS-3 to make a giant leap  
forward in the
automatic analysis of multimedia data, resulting in multiple ``best
performances'' in the international TRECVID benchmark evaluation
for content-based video retrieval. Furthermore, researchers using
DAS-2 and DAS-3 MultimediaN have
earned a ``most visionary research award'' at AAAI 2007
and a `best technical demo award' at ACM Multimedia 2005.

\item The HiRLAM weather forecast
model has been experimented with on wide-area DAS-3.
This model is used by several European meteorological institutes for
their daily weather forecasts. For very high-resolution
forecasts, which will need many processors from multiple  
clusters,
the results are promising.

\end{shortlist}

The main limitation with DAS-3 was that
it was difficult to do large-scale experiments with more clusters
and nodes and to do experiments that need slower (long-haul) networks
and more heterogeneity. For this reason,
the DAS project collaborated with the French Grid'5000 project. The two systems
were connected by a dedicated 10-Gbps light path, aiming to create a
European-scale computer science grid testbed~\cite{bal2007}.
Currently, hardware heterogeneity is being tackled with the
introduction of various HPC accelerators in DAS-4.

 \subsection{FutureGrid}

\minihead{History}
FutureGrid was funded by NSF's Office of Cyberinfrastructure as a result of a proposal submitted in November 2008. It started October 2009 with a four-year budget of \$15M.
    
\minihead{Mission/Vision}
The goal of FutureGrid is to support research on the future of distributed, grid, and cloud computing by building a robustly managed simulation environment or testbed to support the development and early use in science of new technologies at all levels of the software stack: from networking to middleware to scientific applications. The environment will mimic TeraGrid and/or general parallel and distributed systems. This testbed will succeed if it enables major advances in science and engineering through collaborative development of science applications and related software. FutureGrid can be considered as a small science/computer science cloud, but it is more accurately a virtual-machine-based simulation environment.

In many ways, it was conceptually based upon Grid'5000 but is not encumbered by the requirement and responsibility to support production usage. Consequently, FutureGrid is unusual among the infrastructures that we discuss in this \ifreport report. \else chapter. \fi

Although experimental in its early stages, there is a clear trajectory to making FutureGrid a part of the US national cyberinfrastructure.  Specifically it is planned that the FutureGrid research testbed will ``open up'' and become part of XSEDE in fall 2011.

\minihead{Management}
FutureGrid is a partnership of Indiana University (lead, architecture, core software, support), Purdue University (HTC hardware), San Diego Supercomputer Center at the University of California San Diego (monitoring), University of Chicago/Argonne National Laboratory (Nimbus), University of Florida (ViNE, education and outreach), University of Southern California Information Sciences Institute (Pegasus to manage experiments), University of Tennessee Knoxville (benchmarking), University of Texas at Austin/Texas Advanced Computing Center (portal), University of Virginia (OGF, advisory board and allocations), and Center for Information Services and GWT-TUD from Technische Universtit\"{a}t Dresden Germany (VAMPIR). FutureGrid hardware totals about 5,000 cores, located at Indiana, Purdue, Chicago, Florida, Texas, and San Diego. It has a dedicated network (except to Texas) that can be isolated, and it features a programmable network fault generator.
 
In the initial phase, high-level decisions are made by the co-PIs. There are seven working groups covering operations and change management, performance and monitoring, software, system administration \& Networking, Training, Education and Outreach Services, User Requirements and User Support. These groups report biweekly to NSF and the co-PIs. There is a weekly phone call between all collaborators.

Since this is an experimental/research testbed, the focus is less on supporting all users (like the TeraGrid has) and more on specific requirements and understanding the limitation in existing capabilities to support these requirements.  The management structure reflects this design feature.

\minihead{Roadmap/Future}
Formal early use of FutureGrid started in April 2010, and it remained in early usage mode for much of 2010. However, experimental usage has been increasing, with the number of supported projects crossing 25. Standalone production began in November 2010, and FutureGrid is planned to be integrated with XSEDE's other systems in late 2011.
  
\subsubsection{Characteristics}

The system mimics TeraGrid, with a distributed set of conventional clusters as well as systems specific to TeraGrid. Currently the clusters are four IBM iDataPlex systems and a Dell cluster at Texas. There is also a small Cray XT5 and a HTC Condor pool; other specialized systems will be added. Users can request environments that are either VM or bare-metal based with both Linux and Windows.

\subsubsection{Usage Modes}

FutureGrid can be used for developing new applications and software systems probing a variety of interests, including distributed or parallel systems, multicore technologies, and cloud and MapReduce programming paradigms.  Users can request a distributed collection of resources that can be dynamically configured using IBM's xCAT software.  

In general, FutureGrid will allow repeatable system experiments and reliable performance measurements comparing VM and bare-metal environments.

FutureGrid will support both research and education. Early uses re expected to include new computer science and computational science classes that can exploit the special features (e.g., the isolatable network and cloud architecture) of FutureGrid.

\subsubsection{Successes and Limitations}

A major goal and success of FutureGrid is the support of
cyberinfrastructure developers and users who traditionally have not
been major users of TeraGrid/XSEDE. Over half the projects on FutureGrid
have a computer science focus, while computational
biology~\cite{futuregrid1} is the most frequent domain science focus
for the other projects. Project goals cover
interoperability~\cite{futuregrid2} (including standards-based
approaches such as Genesis and SAGA), technology evaluation (e.g., for
adoption of tested technologies by TeraGrid/XSEDE), programming models (e.g., iterative MapReduce),
education~\cite{lsu_scicomp} (with semester-long classes), and computer
science and domain sciences.

The richness and novelty of FutureGrid offerings created unexpectedly
large demands on systems management and user support, leading to
staffing shortfalls. User support for FutureGrid projects is often
end-to-end, not simply issue- or ticket-based; this was reflected in
changes to the user and project-support structure in late
2010. Additionally, the original architecture for FutureGrid was
developed based on initial and predicted use-cases; the actual uptake
has been somewhat different and several original features have not
been exploited, such as the network interrupt capability.

It is ironic that providing a technology aimed at supporting clouds with efficiency of operation and reduced support costs requirements in large data centers itself needs above-average support. The FutureGrid project also did not take advantage of the drastic decrease in disk cost between preparing the FutureGrid proposal and placing system orders, so the FutureGrid systems are underprovisioned in disk space per node.

\section{Commercial Production Distributed Infrastructures\label{sec:cdis}}

\begin{figure}[!h]
  \begin{center}
  \includegraphics[scale=0.77]{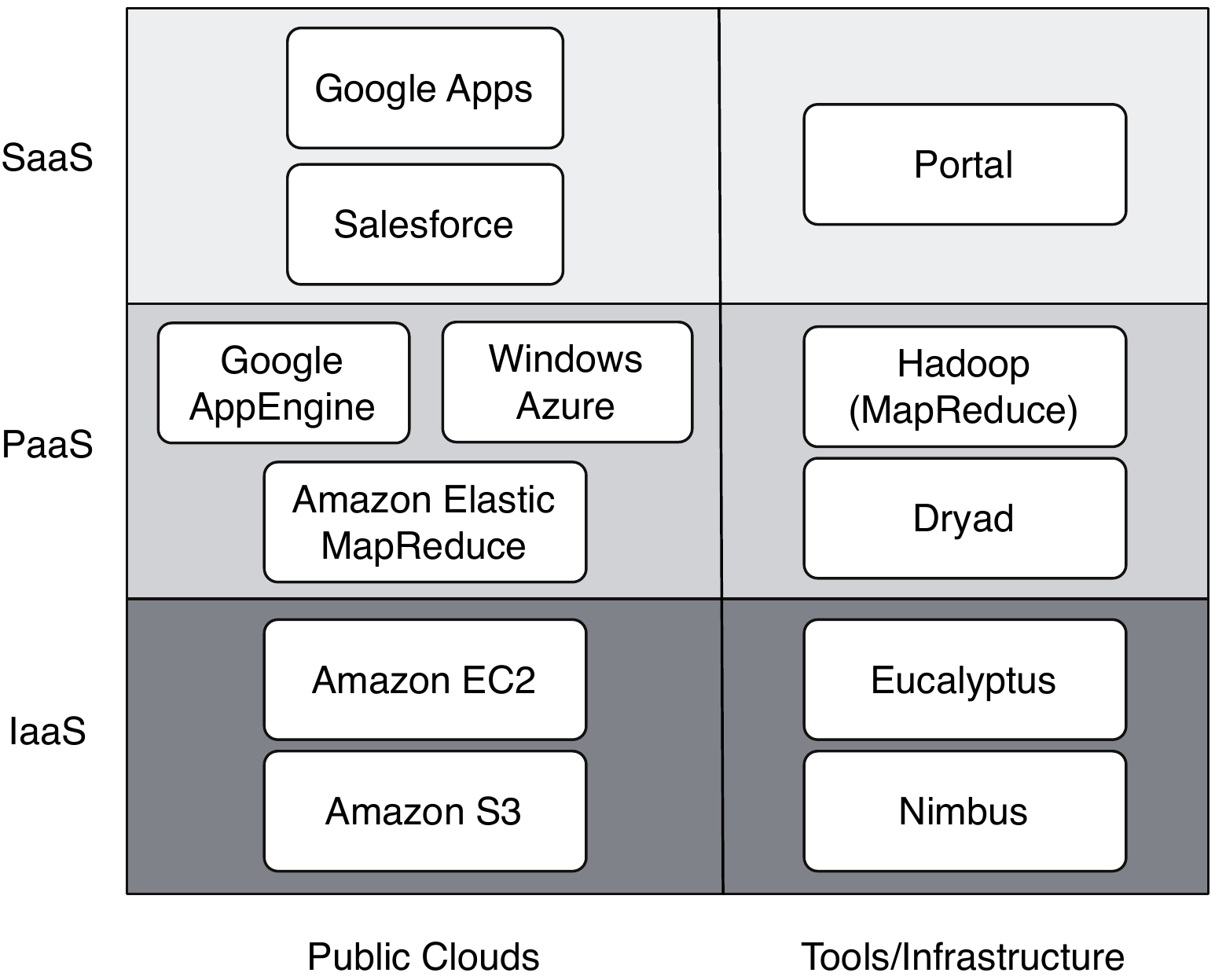}
  \end{center}
  \vskip -0.7 cm
  \caption{Taxonomy of cloud systems, showing the infrastructure-as-a-service, platform-as-a-service, and software-as-a-service layers, as well as examples of both public and private versions of each layer~\cite{cloud_grid_saga}.
\label{fig:cloud-ontology}}
\end{figure}

Currently, most commercial production distributed infrastructures are clouds.
Clouds can be characterized in a number of
ways~\cite{nist_cloud,you08,Armbrust:EECS-2009-28}, including which
layer of services they offer, as shown in
Figure~\ref{fig:cloud-ontology}.  The commonly accepted layers are
infrastructure-as-a-service (IaaS), platform-as-a-service (PaaS), and
software-as-a-service (SaaS).  At each layer, both public and private
clouds can be offered, and each cloud typically uses a set of tools
and infrastructure.  Here, we discuss two public clouds:
an example of IaaS, Amazon Web Services (EC2/S3), and an example
of PaaS, Microsoft Azure. We have
selected these two because they are the examples of commercial
infrastructures on which we are aware that science is being carried out.

Note that the sections that follow describing the commercial infrastructures differ slightly from the previous sections because the goal of these infrastructures is a combination of direct and indirect profit, and neither the management nor the roadmaps for future development are publicly known.

 \subsection{Amazon Web Services}

Amazon's Elastic Compute Cloud (EC2) allows users to rent virtual
computers on which to run their own computer applications. EC2 allows
the deployment of applications by providing a web service through
which a user can boot an Amazon Machine Image to create a virtual
machine, which Amazon calls an ``instance,'' containing any software
desired. A user can create, launch, and terminate server instances as
needed, paying by the hour for active servers, hence the term
``elastic.''  EC2 provides users with control over the geographical
location of instances, which allows for latency optimization and high
levels of redundancy. For example, to minimize downtime, a user can
set up server instances in multiple zones that are insulated from each
other for most causes of failure, such that one backs up the other.

Amazon Simple Storage Service (S3) is a web service that enables users
to store data in the cloud. Users can then download the data or use
the data with other Amazon Web Services (AWS), such as EC2, Amazon Elastic
MapReduce, and Amazon Import/Export. With Amazon S3, a user can charge
others who download data the user makes available. A user can store
up to 5 TB of data in one object but can store as many objects as 
desired. The path to the data is a URL, which makes accessing the
data easy.

\minihead{History}
Amazon announced a limited public beta of EC2 in 2006. Access to EC2 was granted
on a first-come, first-served basis. Amazon added two new instance types (Large
and Extra-Large) in October 2007. Before EC2, Amazon launched S3, its first
publicly available web service, in the United States in March 2006 and in Europe
in November 2007. S3 initially allowed storage of objects up to 5~GB (increased
to 5~TB in December 2010). In May 2008, two more instance types were added,
High-CPU Medium and High-CPU Extra Large. Currently nine types of
instances are available, including Compute-Cluster instances that serve high-end CPU
and interconnect requirements. Compute-Cluster instances use a 10-Gbps
interconnect. Amazon continuously adds features to its portfolio; these features
have included static IP addresses, Availability Zones (specified datacenters), and user-selectable kernels. Amazon added Elastic Block Store (EBS) in August 2008.
EBS allows the user to create storage volumes that can be mounted by EC2
instances. EBS also allows these volumes to be backed-up to S3, providing
persistent storage. EC2 moved from beta to full production in October 2008.

\subsubsection{EC2 Characteristics}

Amazon EC2 presents a virtual computing environment, allowing a user to use web service interfaces to launch instances (virtual machines) with a variety of operating systems, load them with a custom application environment, manage the network's access permissions, and run the image using as many or few systems as desired.

EC2 is intended to have the following characteristics~\cite{ec2-web}:

\begin{shortlist}

\item Elastic: Capacity can be increased or decreased within minutes. A user can commission one to thousands of server instances simultaneously. Because this is all controlled with web service APIs, an application can automatically scale itself up and down depending on its needs.

\item Completely controlled: Users have complete control of their instances. The user has root access to each instance; thus, the user can stop an instance while retaining the data on a boot partition and then subsequently restart the same instance using web service APIs. Instances can be rebooted remotely by using web service APIs.

\item Flexible: The user has the choice of multiple instance types, operating systems, and software packages. Amazon EC2 allows the user to select a configuration of memory, CPU, instance storage, and the boot partition size that is optimal for the choice of operating system and application. Operating systems include numerous Linux distributions, Microsoft Windows Server, and OpenSolaris.

\item Designed for use with other Amazon Web Services: EC2 works in conjunction with S3, Amazon Relational Database Service (Amazon RDS), Amazon SimpleDB, and Amazon Simple Queue Service (Amazon SQS) to provide a complete solution for computing, query processing, and storage across a wide range of applications.

\item Reliable: Amazon EC2 offers a highly reliable environment where replacement instances can be rapidly and predictably commissioned. The service runs within Amazon's proven network infrastructure and datacenters. The Amazon EC2 Service Level Agreement commitment is 99.95\% availability for each Amazon EC2 Region. To ensure a higher-level of availability, users can deploy EC2 instance to different EC2 Regions and Availability Zones. 

\item Secure: Amazon EC2 provides numerous mechanisms for securing the user's compute resources, including customizable firewall settings that control network access to and between groups of instances and isolation of compute instances by using the Virtual Private Cloud (VPC) service.

\item Inexpensive: Amazon EC2 passes on some of financial benefits of Amazon's
scale. Users pay only for the resources that they consume.
\end{shortlist}

EC2 features include Amazon Elastic Block Store (EBS), which offers persistent
storage for Amazon EC2 instances; Amazon CloudWatch, a web service that provides
monitoring for AWS cloud resources; Amazon Virtual Private Cloud (VPC), a set of
isolated compute resources accessible via a Virtual Private Network (VPN)
connection; 
and high-performance computing clusters, tightly coupled computer resources with
high-performance network capability.

\subsubsection{S3 Characteristics}

S3 is based on the idea that Internet storage should be taken for granted. It
is intended to free developers from worrying about how they will store their data,
whether it will be safe and secure, or whether they will have enough storage
available. There are no upfront costs for setting up a storage solution.
The operational costs can be managed by using Amazon's tools and generally depend on the storage usage. However,
depending on the overall amount of storage and usage, the operational costs can
be higher than for an on-premise solution. The functionality of S3 is simple and
robust: store any amount of data while ensuring that
the data will always be available when needed. S3 enables developers to
focus on innovating with data, rather than figuring out how to store it.

A forcing-function for the S3 design was that a single S3 distributed system was needed that  supported the requirements of both internal Amazon applications and external developers of any application. This meant that S3 had to be fast and reliable enough to run Amazon.com's websites, while flexible enough that any developer could use it for any data storage need. S3 was built to fulfill the following design requirements:~\cite{s3-web}
\begin{shortlist}
\item Scalable: S3 can scale in terms of storage, request rate, and users to support an unlimited number of web-scale applications. It uses scale as an advantage: adding nodes to the system increases, not decreases, its availability, speed, throughput, capacity, and robustness.
\item Reliable: Amazon provides different kind of SLAs: The best SLA ensures 99.999999999\% durability with 99.99\% availability. Lower-level SLAs offering, for example, less durability are available. The overall architecture avoids single points of failures. If a failure occurs, the system attempts to repair itself without any downtime.
\item Fast: S3 must be fast enough to support high-performance applications. 
Generally S3 storage can be collocated in the same datacenter as the compute 
instances. Using CloudFront---Amazon's content-delivery network---S3 content can be efficiently distributed.
\item Inexpensive: S3 is inexpensive because it is built from inexpensive commodity hardware components as well as open source software such as Linux and Xen.  It is also hardware-agnostic, so the price decreases as Amazon continues to drive down infrastructure costs.
\item Simple: Building highly scalable, reliable, fast, and inexpensive storage is difficult. S3 offers the user a service that fulfills these properties using an easy-to-use REST-based interface.
\end{shortlist}

We note that these are design requirements, not necessarily operational characteristics.  Amazon has recently had some very public failures that resulted in the unavailability of a large number of applications that depended on AWS ~\cite{aws-failure-2010,aws-failure-2011}.

The S3 architecture is designed to be programming language-neutral, using Amazon's supported interfaces to store and retrieve objects. S3 provides both a REST and a SOAP interface.
Buckets are the
fundamental container in S3 for data storage, and they provide a unique namespace for the
management of objects contained in the bucket. 
Objects (which are stored in buckets) consist of object data and metadata and can range in size from 1~B to 5~TB.
The data portion is opaque to S3, and the metadata is a set of name-value pairs that
describe the object.
Each object is uniquely identified by a key. Together, a bucket name and a key uniquely
identify an object in Amazon S3.

\subsubsection{Usage Modes}

Amazon describes the usage of Amazon Web Services as application
hosting, backup and storage, content delivery,
and so forth~\cite{AWS_case_studies} Most usage of AWS has been built around
the ``usage pattern'' of using EC2 as an infrastructure that is
available on demand, either as a complete substitute for in-house
computing (e.g., hosting web services) or as a resource that can handle
excess demand (e.g, cloudbursting); in both cases S3  is used to
store the required images and data.
In addition, several companies use EC2 or S3 to, in turn, host PaaS-like or SaaS-like capabilities, as described under Successes and Limitations.  In addition,
S3 is often used simply as distributed storage (e.g., for content storage and distribution;
for backup, archiving, and disaster recovery).

Given that a common usage of AWS is as a source of on-demand pool of resources
(spare and instantaneously available), most applications have developed ``glue
code'' that directly spins up instances as needed. Many applications, however,
have made use of other features of AWS, for example, Amazon Queuing Service
(AQS), Elastic Beanstalk, and Elastic MapReduce. In general, many services with
well-defined APIs are emerging that provide easier ways to do more than just stand up an
image instance (a characteristic of IaaS clouds); they extend the
basic IaaS capability to provide SaaS-like capabilities. It is likely that an
increasing number of data analytics services will be provided at this level.

\subsubsection{Successes and Limitations}

Amazon has been successfully used by a wide variety of
users~\cite{AWS_customer_apps}. As of November 2010, Amazon was publicizing 569
customer applications that had been built on top of EC2 and S3. In addition, a
variety of companies are using EC2 and S3 for application hosting, backup and
storage, content delivery, e-commerce, high performance computing, media
hosting, on-demand workforce, search engines, and web hosting. Some well-known
examples of such companies are DropBox, Facebook, Guardian News \& Media,
Playfish, Salesforce.com, and the MathWorks~\cite{AWS_solution_providers,
AWS_case_studies}. Examples of academic and scientific projects using EC2
include the Belle high energy physics experiment's use of the DIRAC framework to
process data using EC2 to supplement existing resources~\cite{belleDIRAC} and
NASA's  use of the Polyphony framework~\cite{polyphony} to execute large workflows
on EC2 in conjunction with existing supercomputers, with Amazon's Simple Queuing
Service used for coordination.

As previously mentioned, Amazon has also had very public failures that primarily impacted public-facing companies, making their products temporarily unavailable~\cite{aws-failure-2010,aws-failure-2011}.  These issues are probably less important for most scientific applications.

On the whole, AWS has been successful and has been a pioneer in the development
of cloud computing; most limitations of AWS are probably limitations of current
virtualization and cloud technology, for example, the limited support for applications
that require tightly coupled parallelism . Currently, the cost of data
movement into and out of AWS is sufficiently expensive that academic
data-intensive applications (e.g., decadal astronomy surveys, bioinformatics
projects that analyze data from next generation sequencers) are unable to
utilize AWS as a production alternative to campus cyberinfrastructure.

 \subsection{Microsoft Azure}

Azure~\cite{win_azure} is an emerging cloud platform developed and operated by Microsoft.  
Azure follows the PaS paradigm, offering an integrated solution for 
managing compute and data-intensive tasks as well as web applications. The platform 
is able to dynamically scale applications without the need to manually manage tasks 
and deployments on the virtual-machine level. In contrast to traditional IaaS clouds 
(e.g., EC2 and S3), Azure provides different benefits: 
first, it operates on a higher level of abstraction and removes the need to 
manage details, such as configuration and patching of the operating 
system; and second, Azure applications are declaratively described and packaged
and are automatically mapped to available hardware by
the fabric controller (which generally
manages the lifecycle of all VMs, monitors them, 
automatically reacts to hardware and software failures, and manages application
upgrades).

\minihead{History}
The foundation for Azure was laid by a memo from Microsoft's chief software architect Ray Ozzie on the Internet service disruption in 2005~\cite{Ozzie:2005fk}. Azure was first announced at the Microsoft Professional Developer Conference 
in October 2008. The initial customer preview included the Azure Storage services: Blob, Queue, and Table storage,
as well as the two kinds of hosted services for web applications and for general compute tasks. Gradually, new features (e.g., support for native code, development tools for Java/PHP, a content delivery
network) have been added to the platform. The latest addition
is a generic Windows VM hosting service. Azure went into production in January 2010.

\subsubsection{Characteristics}

Azure is a group of cloud-related technologies. Parts of these technologies, 
such as the Windows Azure storage and compute services, have 
been specifically designed for cloud environments, while other services are
mainly ports of Microsoft's existing in-house products, for example, SQL Azure from Microsoft's SQL Server.

Windows Azure can be used for different types of on-demand computing and
for hosting generic server-side applications. Azure was designed addressing the following principles:

\begin{shortlist}

\item Simplicity: Azure provides a set of well-defined services that are accessible via standard protocols, such as HTTP.

\item Strong consistency: In contrast to other storage services such as S3, Azure storage provides strong consistency guarantees.

\item Failure tolerance: Failures of the system are  handled by the Azure 
fabric controller, which monitors all applications running in a role environment 
and restarts them if necessary. Each fabric controller is redundantly deployed 
to a cluster of five to seven machines using a master-based replication algorithm. 
Paxos~\cite{279229} is used for master election.

\item Caching: Using standard HTTP header mechanisms such Etag and If-Match HTTP, 
client-side caching of requests is supported.

\item Autonomy: Azure utilizes a hierarchical management structure consisting of 
the fabric controllers and agents that operate according to a set of specified
objectives. 
\end{shortlist}

Azure provides different abstractions as building blocks for creating scalable 
and reliable scientific applications, including web and worker roles for compute services
and blob, table, and queue services for storage.

Windows Azure formalizes different types of virtual machines into
{\em roles}. {\em Web} roles are used to host web applications
and frontend code; {\em worker} roles are well suited for background
processing. While these roles target specific scenarios, they are also
 customizable. Worker roles can, for example, run native code. The
application must implement a defined entry point, which is
then called by Azure. The newest addition are {\em Azure VM} roles, which essentially 
allow users to run Windows Server 2008 VMs on Azure. VM roles give users
more control over the environment than do worker roles but still maintain
PaaS benefits, such as automatic operating system updates,  fault tolerance,
and automatic load balancing. VM roles can also be accessed via the Remote Desktop Protocol and 
are particularly well suited for running more complex applications.

For storing large amounts of data, the Azure storage platform provides 
three key services: \emph{Azure Blob Storage} for storing large 
objects of raw data, \emph{Azure Table Storage} for semi-structured 
data, and \emph{Azure Queue Storage} for implementing message queues.  
The data is storage replicated across multiple data centers to protect 
it against hardware and software failures. In contrast to other cloud 
offerings (e.g., S3), the Azure Storage Services provide strong 
consistency guarantees, i.e., all changes are immediately visible to 
all future calls. While eventual consistency as implemented by S3~\cite{1294281} 
usually offers better performance and scalability, it has some 
disadvantages, mainly caused by the fact that the complexity is moved 
to the application space.

The Blob Storage Service can store files up to 1~TB, which makes 
it particularly well suited for data-intensive applications. 
Further, the access to the blob storage can be optimized 
for certain usage modes.  \emph{Block blobs} can be split into chunks 
that can be uploaded and downloaded separately and in parallel.  
They are well suited for uploading and streaming large 
amounts of data. \emph{Page blobs} manage the storage as an array of 
pages. Each of these pages
can be addressed individually, making page blobs a good tool for
random read/write scenarios. \emph{Azure XDrive} provides a durable
NTFS volume that is backed by a page blob. In particular, legacy
applications that heavily utilize file-based storage can simply be
ported to Azure using XDrive.

The Azure Queue Service provides reliable storage for the delivery 
of messages within distributed applications.  The queue service is 
ideal for orchestrating various components of a distributed application, 
for example, by distributing work packages or collecting results, which 
could be running on Azure or on another resource (e.g., a science cloud).

Azure Table Storage is designed for storing structured data. 
Unlike traditional relational database systems, the table storage is 
designed with respect to scale-out, low cost, and high performance similar 
to Google's BigTable~\cite{1267323} system. For legacy applications,
Azure also provides an SQL server-based relational datastore called 
SQL Azure. In contrast to Azure tables, SQL storage supports common 
relation database features, such as foreign keys, joins, and SQL as the
query language.

\subsubsection{Usage Modes}

Azure provides several core services supporting various application
characteristics and patterns. Compute-intensive tasks naturally map to worker
roles. The communication and coordination between multiple role instances are
commonly done via the Azure storage services or defined communication endpoints.
Worker roles can run either .NET code or native code.

The Azure Queue Service can support batch queue-style operations, namely, the
subsequent execution of a set of tasks. The VMs containing Azure-based
applications can be started on demand or in time to meet a deadline.
More resources can be added at any time to meet a deadline.

Azure resources can be accessed via a user portal as well as by different
command line and GUI utilities, for example, Visual Studio and Eclipse. Many
applications deploy custom portal applications that provide a domain-specific
entry point. Other applications utilize just Azure resources. In some cases,
Azure resources are also used in conjunction with grid/HPC resources, for example, to
offload computation in order to meet a deadline.

Loosely coupled, ensemble-based applications (e.g., parameter sweeps) that
demand a large number of processors but do not require a low-latency
interconnect are particularly well suited for Azure. Workflow-type applications
including applications based on the Windows Workflow foundation can be easily
supported on top of Azure. Data-intensive applications are particularly
well supported. Affinity groups
\ifreport
\else
(discussed in Chapter~\ref{Chap:dataIntense})
\fi
are used as abstraction for supporting the colocation of data storage and
compute instances. On a more fine-grained level, data stored in Azure storage
can be grouped by using a partitioning key. Entities that have the same
partitioning key are guaranteed to be stored on the same server. Further, Azure
provides direct access to a set of public data through the Azure Data
Market~\cite{azure-data-market}.

Scientific problems that do not require high-end HPC hardware and interconnects
and can be easily ported and scaled out on Azure. These applications can benefit
from the ability to acquire and release resources on demand. An increasing
number of applications therefore directly target distributed infra\-structures as
Azure instead of high-end machines. For example, ensemble-based molecular
dynamics approaches utilize multiple sets of simulations of shorter duration
instead of a single, longer simulation to support a more efficient phase-space
sampling. Single ensemble runs spawning up to 8 cores can be encapsulated into
an Azure worker role. Such simulations often need to  acquire
additional resources---for example, if a certain simulation event occurs that
requires the spawning of an additional replica. This type of application can
greatly benefit from Azure's capability to dynamically allocate resources on
demand. For this purpose, Azure provides a Service Management API, which gives
applications a programmatic access for acquiring and releasing resources. This
capability is also useful for
applications where the execution time
and resource requirements cannot be determined exactly in advance, because of
changes in runtime requirements or changes in application structure.

For data-intensive applications Azure provides several interesting 
storage options: xDrive offers file system access to the Azure Storage 
service, which is particularly relevant for applications that manage 
file-based data flows. Blob storage can store large 
amounts of data: a page blob, for example, can store files up to
1~TB.  
Blob storage supports two different data access patterns: block blobs are
designed for continuous access, such as data streaming, while page blobs can
address each of their constituent pages individually and are particularly well
suited for random access. These properties can be mapped to the characteristics
of the respective application; for example, a MapReduce application usually
accesses data in large chunks, which is well supported by the block blob.

\subsubsection{Successes and Limitations}

A number of scientific applications that use worker
roles for compute and/or data-intensive tasks have been ported to Azure.
AzureBlast~\cite{azure_blast},
for example, relies on worker roles for computing bio-sequences. Lately, applications
with more demanding coordination methods have also been ported to Azure; for example, the
Replica-Exchange algorithm has been successfully ported to Azure using the
BigJob framework~\cite{Luckow:2010uq}. The MODISAzure
framework~\cite{Microsoft:2010fk} implements a four-step image pipeline,
including a user portal for analyzing environmental sensor data from
NASA satellites on top of Azure.

Azure imposes scaling limitations. The
largest supported VM has 8 cores, 14~GB of memory and 2~TB of disk space.
Further, MPI applications currently cannot be run on Azure. Whereas other clouds
can run MPI jobs, the performance usually degrades significantly when running
jobs across multiple VMs.

\section{Summary and Conclusions}

Having discussed the infrastructures individually, we now consider
them together, looking at their history and evolution, the
usage modalities they support,
and how their resources are allocated to users,
We conclude the chapter with some 
observations about abstract models and interoperability.

\subsection{The Infrastructures and Their Evolution}

TeraGrid began as an infrastructure to explore grid computing for
compute-intensive tasks, mostly HPC applications, and like DEISA it
became a collection of mostly HPC systems tied together by common
services.  Both OSG and EGEE started as infrastructures to support
data-intensive tasks, where loosely coupled HTC computing could be run
on the distributed datasets.  Although on a smaller scale than 
TeraGrid/DEISA or EGEE/EGI/OSG, the NGS initially focused on both
data-intensive computing (HTC) and HPC.

Most of the research infrastructures (Grid'5000, PlanetLab, DAS) were
bottom-up developments that grew out of computer science research
needs; they were collaborations of groups of computer scientists who
realized their research would benefit from larger-scale platforms that
could be developed and supported only by such collaborations.
FutureGrid, on the other hand, was a top-down project, which came from
the US NSF deciding to build and support a grid for such
research and issuing a call for proposals.

The commercial infrastructures appear to have dual motivations, though
understanding the internal decisions within the corporations that have
built them is not easy.  EC2 and S3 are widely thought to have been an
effort by Amazon to sell spare capacity, as the company's own operations
require its peak capacity only for short periods each year.  Azure
has been developed by Microsoft as a way to adapt to a new business model
comprising advertising-supported services and software, with the
expectation that this model will lead to increased revenue.

Some of the technological advances and economic trends behind EC2/S3
and Azure, and cloud computing in general, relate to advantages
arising from the economies of scale achieved by large data centers:
the lowering of data center energy and management costs along with the
increasing scale and efficiency of operation. Others arise from
requirements such as aggregation and dealing with large volumes of
datasets or from the energy costs of data movement.  In general, the
rise of the datacenter to support web-scale computing
requirements has been an important driver for the recent advances in
cloud computing.

\subsubsection{Evolution and Supported Capabilities}

Understanding the evolution of certain infrastructure capabilities in
response to application and user needs is both instructive and
interesting. Given OSG's need to support HTC, Condor has evolved from
a scavenging system in the late 1980s to become the basic building block for OSG's
infrastructure in the 2000s. Condor Flocking, which provides
aggregation of resources, is a fine example of continuous transition
versus discontinuous transition.  Similarly, experiences from
SETI@home led to BOINC, which was then used for other @home
applications, such as
climate{\em prediction}.net.

Gateways on TeraGrid emerged when a number of computationally savvy
application developers realized that simplifying the process for
using TeraGrid resources (identification and authorization of the user as well
as the submission mechanisms for the work to be done) would allow other people in their communities
to benefit from those resources.  The
gateways that have been developed often use a graphical user interface
to hide complexity, and provide capabilities such as workflows,
visualization software and hardware, resource discovery, job execution
services, access to data collections, applications, and data analysis
and movement tools.  The number of cycles used through science
gateways increased by a factor of 5 from 2007 to 2008.  By working
with some of the initial gateways developers, TeraGrid has
developed capabilities that can be used by other developers to build
new gateways.

However, in several cases the requirements of
a class of distributed applications are often out of phase with the deployed
capabilities of infrastructure.
One example is the requirement of
distributed pilot job applications~\cite{novel_submission_mode} to
simultaneously use multiple resources on production grids to obtain
results more quickly by using a coscheduling capability.
This is an interesting case study because it involves both policy and
technical challenges. The policy issues have been a barrier because
HPC centers are unwilling to relinquish the batch-queue mode of
operation on individual systems.  Technically, while methods other
than coscheduling can clearly meet this requirement,
such as statistical/probabilistic approaches to co-allocation or
best-effort co-allocation, these have not been made available on
production resources.  The emphasis on batch-queue mode, corresponding
to an emphasis on overall utilization of a HPC resource, has inhibited
other modes of computing, such as urgent computing, ensembles, and
quality-of-service-based  (QoS-based) computing (e.g., user {\it x} will be allowed {\it y} jobs
over period {\it z}).

Another example of a new type of application is found in dynamic
data-driven distributed application systems (DDDAS).
The growth of DDDAS applications has been driven by the emergent abundance of accessible sensor data and the
desirability of coupling real-time simulations to live sensor data,
combined with the maturity of workflow tools. Currently, none of the science infrastructures
can support large-scale DDDAS out of the box and without significant
customization.  For applications such as
LEAD~\cite{lead} and SCOOP~\cite{scoop}
there is a need for guaranteed throughput, which could
be supported by coscheduling, high-priority mechanisms, or QoS-based
computing, none of which are generally available.  Beyond this, OSG
and EGEE/EGI support HTC but not large-scale HPC, while TeraGrid, DEISA,
and NGS support HPC but do not natively support dynamic requirements.

Other external factors will cause new types of distributed
applications to come of age. Anticipating these trends and
supporting them on science infrastructures would benefit
the wider scientific community.  As new types of applications
appear, the underlying infrastructure and capabilities also
change, often more quickly than the timescale on which previously
developed scientific distributed applications were expected to remain
usable.  For example, clouds have rather suddenly emerged and become
prominent.  However, the basic principles and requirements of
distribution have not changed; the fundamental problem of coordinating
distributed data and computation remains. Therefore, it is imperative
that distributed application developers consider developing their
applications using programming systems, tools, and interfaces that
provide immunity from the natural evolution of infrastructure and
capabilities.  Well-designed distributed programming abstractions can
be critical in supporting these
requirements~\cite{grid_cloud_interop}.

\subsection{Usage Modalities Supported}

Usage modalities can be classified as user-intent,
when-to-run,
submission-mech\-anism, targeted resources, and
job/resource coupling modalities.
In this subsection, we discuss each, including which infrastructures support
them.

\subsubsection{User-Intent}

The user-intent modalities are production,
exploration/porting, and education.  Of the infrastructures we have
examined, all the science infrastructures (TeraGrid/XSEDE, DEISA, OSG, EGEE/EGI,
and NGS) support all of the user-intent modalities.  The research
infrastructures (Grid'5000, PlanetLab, DAS, FutureGrid) generally
do not support science production, although they do support computer
science experiments.  They also support exploration/porting and
education.  The commercial infrastructures (AWS, Azure) support all
three user-intent modalities, but these modalities generally are not considered 
separately; rather, they are all just usage, and the intent
of usage is not the concern of the commercial infrastructures.

\subsubsection{When-to-Run}

When-to-run modalities include batch, interactive, urgent
(immediate), urgent (high-priority), and reservation.   Batch
is not the primary usage mode on clouds, but it can easily be supported on clouds.
For example, Azure queues can be used simply as submission
queues for worker roles. The interactive modality is supported
on the commercial infrastructures.  On some TeraGrid/XSEDE resources, it is supported
when prearranged with the resource owner.
On DEISA, it is supported only for setup, test, and development, not for
production.  On visualization resources within TeraGrid/XSEDE and NGS, it is
supported.  Note that in most cases, a clever job (such as a shell)
submitted to a batch
queue can support an interactive session.  The research infrastructures
all support interactive usage, although on DAS it is (by default) limited to 15 minutes
during the daytime to allow quick access to a large portion of the resources.
In some situations, some TeraGrid/XSEDE resources support urgent usage and reservations,
as do OSG and EGEE/EGI, in all cases subject to advance discussion with the
infrastructure.  DEISA and NGS do not support urgent usage, though NGS
does support reservations, again under some circumstances.
Of the research infrastructures, Grid'5000 and DAS can support urgent
usage and reservations in some situations, PlanetLab does support
urgent usage in general but has limited support for reservations, and FutureGrid
does not yet have a determined policy on urgent computing or reservations.
The ideas of urgent computing and reservations
are not directly supported on the research infrastructures, but the basic ideas
can be supported by clever use of applications.

\subsubsection{Submission-Mechanism}

Four submission-mechanism modalities exist: command lines, grid
tools, science gateways, and metaschedulers.  In science
infrastructures, TeraGrid and DEISA support the first three, and
XSEDE aims to develop metascheduling, which exists under some tools
for single-processor jobs.  OSG does not allow user login to compute
nodes and therefore does not allow command-line submission, but it
does support the other three modalities.  EGEE/EGI, while generally a
partner with OSG, supports all four modalities.  NGS also supports all
four modalities.  Of the research infrastructures, Grid'5000 supports
the first three modalities, while DAS supports command-line and grid
tool submission and is experimenting with metaschedulers.  PlanetLab
supports only the command-line modality, but users can add other
layers once they have the resources.  It is not yet clear which of
these FutureGrid will support.  Of the commercial infra\-structures,
EC2, similarly to PlanetLab, allows users to manage resources using a
web portal, a command-line client, and various other client
applications. Having started a resource, users can log into these
resources using SSH or the remote desktop protocol for Windows
resources. Similarly, Azure provides a portal application for managing
resources and deployment of applications. In addition, Azure resources
can be managed from within Visual Studio and Eclipse. Direct access to
resources is possible by using the Remote Desktop Protocol.

\subsubsection{Targeted Resources}

Targeted resource modalities include the coupling of multiple
resources of the same type within the infrastructure, multiple types
of resources within the infrastructure, and the coupling of these resources
with other resources that are not part of the infrastructure.
To understand these, we also need to know what types of
resources are in the infrastructure and whether there are
infrastructure-wide policies that support tools and services to
enable the coupling, either concurrent or sequential, of multiple
distinct resources, even if they are not coupled by the resource
providers.

TeraGrid resources included HPC, HTC, storage, and data analysis and
visualization resources.  On TeraGrid, one could use multiple TeraGrid
resources together. One could also, with a fair amount of work, use
TeraGrid resources and non-TeraGrid resources together; however, with
the exception of science gateways, this was not directly supported by
TeraGrid.  DEISA resources are strictly HPC and storage resources. On
DEISA, the use of multiple DEISA resources together is a supported
usage modality, but the use of DEISA resources with other resources is
not supported.  OSG resources are primarily HTC and storage resources,
which can be used together and can also be used with HTC resources
from other infrastructures.  EGEE (and EGI) and NGS are also primarily
HTC and storage resources.  They are designed to be used together and
with other standards-compliant resources.  The research
infrastructures (Grid'5000, PlanetLab, DAS, and FutureGrid) are
designed primarily for coordinated use of the resources within their
own infrastructure. This does not mean that they cannot be used with
other resources, only that this is not the primary concern of the
infrastructure developers. We note, however, that FutureGrid is
specifically designed to use standards-compliance to allow external
resources to be used together with internal resources.  EC2 and Azure
resources can easily be combined with other types of resources, such
as grid resources, using tools and capabilities such as SAGA~\cite{saga}.

In general, several factors influence the use of
resources.  Where there are a small number of resources with large
individual capacity (e.g., TeraGrid, DEISA), there is less incentive, and
perhaps less user need, to use multiple resources together.
In many of the current infrastructures, it is also more difficult to
use multiple resources together than to use a single resource.
Similarly, where an infrastructure has a large capacity internally,
there can be less incentive and less user need to use this infrastructure with
resources from another infrastructure. Furthermore, using multiple
infrastructures together inevitably involves extra work. In both cases,
nontechnical issues also are at play, such as the incentive of the
resource owners or infrastructure partners to work with other resource
owners or infrastructures, who may see advantages in having a captive
market, may have to support multiple sets of users with
different expectations and requirements, or may feel as though
they are competing against the others.

Our analysis reveals a spectrum of infrastructure
types.  At one end of the spectrum is a small number (O(10)) of
large resources, such as TeraGrid and DEISA.  In the middle of the spectrum is a
moderate number (O(100)) of smaller resources, such as OSG and EGEE/EGI.
And at the far end of the spectrum is a large number (O(10000+)) of small
resources, such as volunteer computing grids. Unsurprisingly,
most infrastructures are built around roughly ``equal'' styles and types of
resources, and so there remains a challenge for applications or
users that might want or need to span different infrastructures.

\subsubsection{Job/Resource Coupling}

The job/resource coupling modalities are independent,
independent but related, tightly coupled, and dependent.  (Note: an
infrastructure might support the running of MPI jobs on a particular
resource within that infrastructure, but tightly coupled is used here
in the distributed context, meaning across multiple resources.)
TeraGrid, NGS, Grid'5000, and DAS support all four, as will
FutureGrid.  DEISA, OSG, and EGEE/EGI support all but the tightly coupled
modality.  PlanetLab and EC2 support none of these; they provide resource
slices and resources respectively, which the user can then use as
desired.  Azure supports all four, with the limitation that tightly
coupled jobs are best when the VM is constrained to a node/processor,
and MPI jobs in particular are supported only on a single VM instance,
not across multiple instances, because of a limitation of the communication
endpoint model that is used, which does not support dynamic port
ranges.

\subsection{Allocations and Usage}

The methods for obtaining the ability to use resources on the
infrastructures also vary.  Four basic paradigms exist.  In all
cases, the infrastructure owners have some process for deciding who is
eligible to use the resources.  For example, TeraGrid/XSEDE can be used
by researchers led by a person affiliated with a US institution who
intends to do research that will be published in the open
literature. Similarly, DAS can be used by researchers within or
collaborating with the five organizations that own and host the DAS
resources.

In the first paradigm, as on TeraGrid/XSEDE, DEISA, and FutureGrid, individual users
write proposals (that may be for themselves or a team) for resources,
and these proposals are peer reviewed.  On TeraGrid/XSEDE, a proposal can
also represent a community account, such as for a science gateway,
where the proposer will reallocate the resources among a community.
These proposals effectively return an allocation of the resources over
a period of time.  For both TeraGrid/XSEDE and DEISA, allocation
decisions are made by the project.  FutureGrid currently
uses a review by the FutureGrid project to provide access to the grid;
but as FutureGrid becomes a production element in XSEDE, this
process will be incorporated in XSEDE's regular review process.
Once the allocation decisions are made, a queuing system is used
on most resources, where users submit jobs and the system maps the
queued jobs to the resources over time.

In the second paradigm, as on OSG and EGEE/EGI, decisions about
which users can use which resources are made by the resource owners,
in contrast to the central decisions made in the first paradigm.
On OSG, the resource owners
generally reserve some fraction for their own use and offer unused
resources to others in one or more virtual organizations (VOs).  EGEE/EGI
resource owners simply offer their resources to one or more VOs.  All
users are members of at least one VO, and through their VO they
have the opportunity to compete for use of the resources where their
VO is able to run.

In the third paradigm, as on the NGS, PlanetLab, Grid'5000, and DAS,
no process exists for allocating the resources,
and all the users fight for them through batch queues or other
mechanisms, possibly with first-come, first-served or fair-share
policies.

In the fourth paradigm, in use on the commercial grids EC2/S3 and Azure,
usage is simply paid for.  There are no batch queues; when a user requests
resources, they are either available or not.

\subsection{Applications Use of Infrastructures}

Our discussion of applications and infrastructures and our own
experience in developing applications for both parallel and
distributed infrastructures point to certain barriers in the
effective development and deployment of distributed applications.  When developing an
application, the developer has to frame the potential application in
terms of functions that can be implemented on the infrastructure on
which the application will run.  In parallel computing, there has been
an approximately 20-year span under which the abstract infrastructure
model has been well known: a set of interconnected nodes, each with a
processor and a memory.  The MPI standard assumes this model.
As multiprocessor nodes and multicore
processors have appeared, however, this model
is no longer sufficient to write
optimal code, though it is still sufficient to write portable code.

For distributed applications, however, no abstract model of the
potential infrastructures seems sufficient. Not only do all the
hardware and system level issues that challenge a parallel program
developer also challenge a distributed application developer, but one
can argue that issues of policy, deployment, and execution tools, and
environment make the distributed applications more complicated.
Additionally, the lack of an abstract model of potential
infrastructures is coupled with an empirical observation that similar
``functionality'' has been provided by using very different tools and
capabilities. For example, two of the most popular and large
high-throughput distributed-computing infrastructures---OSG and
EGEE/EGI---have very different
environments for data management and managing jobs or tasks, thus creating
a barrier to interoperability.

Developers who use a model of a volunteer computing grid and want to run on
some of the DEISA systems are not making good use of the systems
and will likely not successfully pass through the review process to obtain an
allocation to run on such systems.  On the other hand, an application written
to run well on DEISA probably will not run at all on a volunteer
computing grid.  Additionally, there is no equivalent to the MPI
standard for distributed computing, although of course it would be 
hard to have such a standard without first having a common abstract
infrastructure model on which to think about and design a standard.

Infrastructure providers have a similar problem.  They need to design and provide
an infrastructure that meets a set of user needs, so that users can build applications
that run on the infrastructure.  But users generally state their needs  in terms of what
they think is feasible: what they think the infrastructures can provide.  In some cases,
the providers and the users can work things out. For example, the EGEE/EGI and
OSG infrastructures have been driven by a specific model of mostly sequential
jobs, originally coming from the high energy physics community.  These infrastructures
providers have been able to
build an infrastructure that meets this need, and application developers in
other science domains have
used this model and built new applications that work.

Perhaps the answer is that there is no single abstract
infrastructure model for distributed applications, but rather there
are a number of distinct models, and application developers need to
choose one of them and then use the infrastructures that match their
model.  If this is so, there could be a standard for each model,
similar to the MPI standard that has been used for the model of
parallel nodes, each with CPUs and associated memory.  In some ways,
EGEE/EGI and OSG use a model similar to this, one of distributed
slots of computing, each with some associated storage.  But TeraGrid
has an implicit variety of models, one of which is the distributed set
of parallel computers that is the main model in DEISA.

In general, some standards are important in all the infrastructures
we have discussed.  For example, GridFTP is supported by all the
infrastructures.  In other areas, standards are used in some of the
infrastructures, particularly where they provide a needed capability.
For example, many OGF standards are supported by EGEE/EGI because this is
really a federated infrastructure, where different providers choose
different software on different parts of the infrastructure.
Standards allow these different choices to work together.  On the
other hand, TeraGrid does not use many of the OGF standards.  Instead,
the project requires all parts of the infrastructure to use the Globus
Toolkit, which becomes a de facto standard.  This requirement obviously leads to
difficulties if a user wants to use EGEE/EGI and TeraGrid together; but
because there are not many such users, they can deal with this
situation by writing custom adaptors or using tools that have already
developed adaptors, such as SAGA-based tools or AHE~\cite{ahe}.

A final issue for the use of science grid infrastructures is 
the timescale of change.  Currently, the infrastructures are
changing faster than the applications. This situation is partly because
distributed infrastructures and their provisioning are
correlated to existing and emerging technologies; distributed
applications are not easy to reformulate or refactor. For example,
infrastructures generally appear to last for three to seven years.  But
applications often take years of development and then are expected to
last for 20 or more years.

Currently, there appears to be no satisfying solution to this discrepancy,
but perhaps the use of a small number of distributed
abstractions that enable the decoupling of applications from
infrastructures will help.  For example, given that a large number of
applications now use MapReduce, infrastructure providers will
likely continue to support this abstraction as they change their
infrastructure.  And thus there will emerge MapReduce the pattern,
MapReduce the programming model and execution environment, and finally
specific implementations of MapReduce on different infrastructures.
Identifying such abstractions is one of the goals of the book from which
this report is derived.

\newpage
\appendix

\bibliographystyle{plain}
\bibliography{dpa_book}
\end{document}